\documentclass[usenatbib]{mn2e}

\usepackage{amsmath}
\usepackage{epsfig}
\usepackage{txfonts}
\usepackage{pict2e}
\usepackage{color}

\newcommand{\Jsurv}{\mathcal{P}_{\rm s}}
\newcommand{\nuin}{\nu_{\rm i}}
\newcommand{\zin}{z_{\rm i}}
\newcommand{\xin}{x_{\rm i}}
\newcommand{\Tin}{T_{\rm i}}
\newcommand{\Teq}{T^{\rm eq}_{\rm e}}
\newcommand{\xp}{x_{\rm p}}

\newcommand{\eV}{{\rm eV}}

\newcommand{\Kel}{{\rm K}}

\newcommand{\cm}{{\rm cm}}

\newcommand{\GHz}{{\rm GHz}}

\newcommand{\expf}[1]{{{\rm e}^{#1}}}
\newcommand{\Jbb}{\mathcal{J}}

\newcommand{\zmu}{{z_{\mu}}}

\newcommand{\ye}{{y_{\rm e}}}

\newcommand{\xe}{x_{\rm e}}

\newcommand{\xc}{x_{\rm c}}

\newcommand{\id}{{\,\rm d}}

\newcommand{\beq}{\begin{equation}}   %

\newcommand{\eeq}{\end{equation}}   %

\newcommand{\beqa}{\begin{eqnarray}}   %

\newcommand{\eeqa}{\end{eqnarray}}   %

\newcommand{\beal}{\begin{align}}
\newcommand{\enal}{\end{align}}

\newcommand{\bspl}{\begin{split}}

\newcommand{\espl}{\end{split}}

\newcommand{\bsub}{\begin{subequations}}

\newcommand{\esub}{\end{subequations}}

\newcommand{\bmulti}{\begin{multline}}   %

\newcommand{\beqm}{\begin{mathletters}}   %

\newcommand{\eeqm}{\end{mathletters}}   %

\newcommand{\me}{m_{\rm e}}

\newcommand{\Ne}{N_{\rm e}}

\newcommand{\Te}{T_{\rm e}}

\newcommand{\Tg}{T_{\gamma}}

\newcommand{\The}{\theta_{\rm e}}

\newcommand{\Thg}{\theta_{\gamma}}

\newcommand{\sigT}{\sigma_{\rm T}}

\newcommand{\pot}[2]{#1 \times 10^{#2}}

\newcommand{\url}[1]{{\tt #1}}

\usepackage{grffile}
\usepackage{graphics}
\usepackage{booktabs}

\title[Photon injection]
{Green's function of the cosmological thermalization problem II: effect of photon injection and constraints}

\author[Chluba]{Jens~Chluba$^{1, 2}$\thanks{E-mail:jchluba@ast.cam.ac.uk}
\\
$^{1}$ Kavli Institute for Cosmology Cambridge, Madingley Road, Cambridge, CB3 0HA, UK
\\
$^{2}$ Department of Physics and Astronomy, Johns Hopkins University, 
3400 N. Charles St, Baltimore, MD 21218, USA
}

\begin{document}

\date{{\vspace{-1mm}Accepted 2015 September 24. Received 2015 June 28}}

\maketitle

\begin{abstract}
The energy spectrum of the cosmic microwave background (CMB) provides a powerful tool for constraining standard and non-standard physics in the primordial Universe. Previous studies mainly highlight spectral distortions ($\mu$-, $y$- and $r$-type) created by episodes of early energy release; however, several processes also lead to copious photon production, which requires a different treatment. Here, we carry out a first detailed study for the evolution of distortions caused by {\it photon injection} at different energies in the CMB bands. We provide detailed analytical and numerical calculations illustrating the rich phenomenology of the associated distortion signals. 
We show that photon injection at very high and very low frequencies creates distortions that are similar to those from pure energy release. In the $\mu$-era ($z\gtrsim \pot{3}{5}$), a {\it positive} or {\it negative} chemical potential can be formed, depending on the balance between added photon energy and number. At lower redshifts ($z\lesssim \pot{3}{5}$), partial information about the photon injection process (i.e., injection time and energy) can still be recovered, with the distortion being found in a partially Comptonized state. 
We briefly discuss current and future constraints on scenarios with photon production. We also argue that more detailed calculations for different scenarios with photon injection may be required to assess in which regimes these can be distinguished from pure energy release processes.
\end{abstract}


\section{Introduction}
\label{sec:Intro}
The immense potential of cosmic microwave background (CMB) spectral distortions as a unique probe of standard and non-standard physics in the early Universe has spurred renewed theoretical activity \citep[see][for overview]{Chluba2011therm, Sunyaev2013, Chluba2013fore, Tashiro2014}. Commonly discussed spectral distortions are primarily due to energy release \citep{Zeldovich1969, Sunyaev1970mu, Burigana1991, Hu1993}, for example, because of decaying relic particles \citep{Sarkar1984, Kawasaki1986, Hu1993b, Chluba2013PCA} or the damping of primordial acoustic waves at small scales \citep{Hu1994, Chluba2012, Chluba2012inflaton}, which causes $\mu$-, $y$- and $r$-distortions. In the future, these signals may be detected with an experiment similar to PIXIE \citep{Kogut2011}, a satellite concept which could provide a major leap forward in sensitivity over COBE/FIRAS \citep{Fixsen1996} and may open a new unexplored window to the early Universe.

While most theoretical studies focused on distortions caused by early energy release, spectral distortions can also be created by direct {\it photon injection}. Physically, the difference is that initially {\it only} the photon field is affected, whereas for pure energy release it is assumed that the baryons and electrons are heated but no photons are directly added. For photon injection, part of the injected photon energy is converted to heat (through Compton scattering and photon absorption), while the other part remains in the scattered initial photon spectrum. 

One inevitable distortion created by photon injection is caused by the cosmological recombination process through uncompensated atomic transitions in the hydrogen and helium plasma around $z\simeq 10^3$ \citep[see][for overview]{Sunyaev2009}. At this late stage, Comptonization is already very inefficient so that the emitted photons just redshift due to the expansion of the Universe \citep[e.g.,][]{Jose2006, Chluba2006b}. The shape of the cosmological recombination spectrum thus directly reflects the dynamics of the recombination process. However, for photon injection at earlier stages ($z\gtrsim 10^4$), Compton scattering redistributes photons more strongly over energy, so that direct information of the initial photon injection process is  (partially) erased. In addition, at low frequencies, double Compton (DC) and Bremsstrahlung (BR) emission/absorption become more efficient, so that detailed calculations are required to interpret the signals.  

Other examples associated with photon injection/destruction at a wide range of energies are related to {\it photon-axion conversion} in the presence of primordial magnetic fields \citep{Tashiro2013, Ejlli2013}, {\it graviton-photon conversion} \citep{Dolgov2013}, {\it non-evaporating spinning black holes} \citep{Pani2013}, {\it superconducting strings} \citep{Ostriker1987, Vilenkin1988, Tashiro2012}, {\it gravitino dark matter} \citep{Lamon2006} or simply {\it primary} and {\it secondary photons} created in particle cascades from decaying or annihilating relics \citep{Shull1985, Chen2004, Slatyer2009, Huetsi2009, Valdes2010, Slatyer2015}. 

For most previous estimates related to CMB spectral distortions, the simplifying assumption that photon injection can be treated similar to energy release was applied; however, as we explain here, this approximation is only justified in certain regimes of injection redshifts, $\zin$, and frequencies, $\nuin$. For instance, for injection at very low energies, $h\nuin\lesssim 10^{-4} k\Tg$, where $\Tg(z)$ is the temperature of the CMB blackbody at redshift $z$, photons are efficiently absorbed and converted into heat by the BR and DC processes before Comptonizing. Similarly, at high frequencies, during the pre-recombination era ($z\gtrsim 10^3$), most of the photon energy simply leads to heating of the medium through electron recoil. However, at intermediated frequencies the problem is richer.

Here, we carry out a first systematic study for the evolution of distortions created by photon injection\footnote{Photon destruction can be treated as {\it negative} photon injection, and thus is also covered by our treatment.}. We illustrate the rich phenomenology of the resulting distortion signal for single photon injection, providing both numerical results and analytical approximations and estimates. In particular, we introduce the {\it photon injection Green's function} (Sect.~\ref{sec:Greens_nu}), which allows computing the distortions for more general photon injection scenario. We also briefly discuss constraints on single photon injection scenarios  (Sect.~\ref{sec:FIRAS_const}) derived from COBE/FIRAS \citep{Fixsen1996} and big bang nucleosynthesis (BBN) constraints \citep{Simha2008}.

For the main calculations performed in this paper, we assume that photons are only injected in the CMB band, $h\nu\lesssim 30 k\Tg$. In particular, we restrict ourselves to energies below the pair-production threshold, above which significantly more processes become relevant \citep[e.g.,][]{Svensson1984, Zdziarski1989} also related to interactions with light elements \citep[e.g.,][]{Kawasaki2005}. In Sect.~\ref{sec:gamma_injection}, we briefly discuss photon injection at higher energies, for which most of the energy is eventually converted into heat. However, since several soft secondary photons can be produced more detailed calculations seem to be required. For similar reasons, we highlight the necessity to perform more detailed calculations for different scenarios associated with photon injection to assess whether these can be distinguished from pure energy release.

\section{Green's function for photon injection}
\label{sec:Greens_nu}
For small spectral distortions, the photon Boltzmann equation can be linearized. The thermalization problem can thus be solved using a Green's function method. The Green's function for pure energy injection, $G_{\rm th}(\nu, z)$, was already discussed earlier \citep{Chluba2013Green}. 
The final spectral distortion caused by different energy release histories can thus be computed as  \citep{Chluba2013Green}
\beal
\Delta I_\nu(z=0)=\int G_{\rm th}(\nu, z') \,\frac{\id(Q/\rho_\gamma)}{\id z'}\id z'.
\end{align}
Here $\id(Q/\rho_\gamma)/\id z'$ parametrizes the energy release history. The energy release Green's function is well approximated as a sum of $\mu$ and $y$ distortion with an additional temperature shift \citep{Chluba2013Green}. For higher accuracy, the $r$-type (non-$\mu$/non-$y$) distortion \citep{Chluba2011therm, Khatri2012mix, Chluba2013Green} becomes important in the $\mu$-$y$ transition era ($10^4\lesssim z \lesssim \pot{3}{5}$) and in principle allows distinguishing between different energy release scenarios \citep{Chluba2013fore, Chluba2013PCA}.

In contrast, here we discuss photon injection at different injection frequency $\nuin$ and redshift $\zin$. Assuming that the injected photon number is small, we can again compute the final distortion using a Green's function approach, but this time we have to perform computations on a grid of injection frequencies and redshifts. This allows accelerating the computation of the resulting distortion for more general photon injection histories and also provides detailed insights that can be used to develop analytic approximations.

In practice, we simply add photons to the photon distribution at $\zin$ using a narrow Gaussian centered at $\nuin$ and then compute the final spectral distortion at $z=0$ using {\tt CosmoTherm}\footnote{\url{www.Chluba.de/CosmoTherm}} \citep{Chluba2011therm}. With this procedure we obtain the photon injection Green's function, $G_{\rm in}(\nu, \nu', z')$, which describes the spectral response for different injection frequencies $\nu'$ and redshifts $z'$. This Green's function in principle has two parts, one that is similar to the energy release Green's function and the other that arises directly from the scattered photon distribution; however, from the numerical point of view it is easier to include both parts simultaneously.
The spectral distortion can then be computed as
\beal
\label{eq:Greens_injection}
\Delta I_\nu(z=0)=\int G_{\rm in}(\nu, \nu', z') \,\frac{S(\nu', z')}{h\nu'} \id \nu' \id z',
\end{align}
where $S(\nu, z)$ is the photon production or source term. We define it such that at any moment the relative changes of the photon number and energy densities are $\Delta N_\gamma(z)/N_\gamma(z)=\int [S(\nu, z)/h\nu]\id \nu$ and $\Delta \rho_\gamma(z)/\rho_\gamma(z)=[\alpha_\rho/k\Tg(z)] \int S(\nu, z)\id \nu$, respectively. Here, we set $\alpha_\rho=G_2^{\rm Pl}/G_3^{\rm Pl}\approx 0.3702$ with $G_k^{\rm Pl}=\int x^k/(\expf{x}-1)\id x=k!\zeta(k+1)$. The Green's function is thus normalized as
\beal
\label{eq:G_norm}
\frac{4\pi}{c} \int G_{\rm in}(\nu, \nu', z') \id \nu=\frac{h\nu'}{k\Tg(z')}\,\alpha_\rho\, \rho_\gamma(0)
\equiv \frac{h\nu'}{1+z'}\, N_\gamma(0).
\end{align}
Below we discuss the dependence of $G_{\rm in}(\nu, \nu', z)$ on the injection frequency and redshift in the different distortion eras. We furthermore give simple analytic approximations that can be used to compress the information needed to describe the Green's function for a range of scenarios, thereby accelerating the computation. 

\subsection{Basic ingredients and assumptions}

\subsubsection{Energy exchange by Compton scattering}
\begin{figure}
\centering 
\includegraphics[width=1.02\columnwidth]{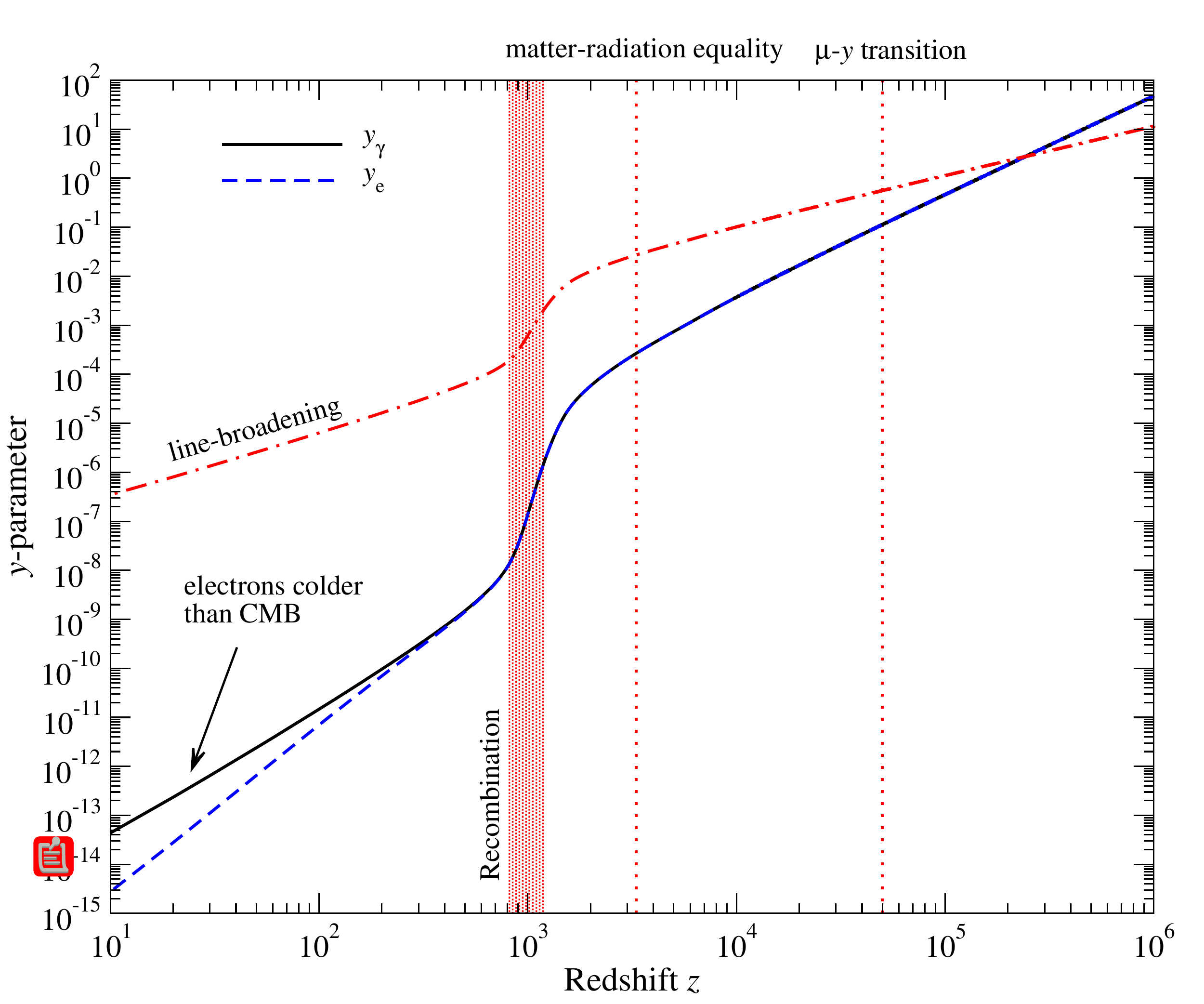}
\caption{Dependence of the $y$-parameters, $y_\gamma$ and $\ye$, on redshift. After recombination, the $y$-parameters drop strongly since the number of free electrons decreases exponentially. At late times, electrons drop out of equilibrium with the photons so that $\ye<y_\gamma$. Around $z_{\rm K}=\pot{5}{4}$, we have $\ye\simeq y_\gamma\simeq 0.1$. The strength of line-broadening, $\Delta \nu / \nu \simeq 2 \sqrt{ y_\gamma \ln 2}$, is also illustrated. After recombination, it becomes much smaller than $\Delta \nu / \nu\simeq 10^{-3}$ and can be neglected for most applications. Photon drift towards lower frequencies through electron recoil is roughly given by $\Delta \nu/\nu\simeq - (h\nu/k\Tg) \, y_\gamma$.}
\label{fig:y-parameters}
\end{figure}

The energy exchange between electrons and photons is controlled by Compton scattering, which can be described using the Kompaneets equation \citep{Kompa56}. The total Compton-$y$ parameter for the evolution of the photon distribution between redshift $z$ and today is given by
\beal
\label{eq:y_gamma}
y_\gamma =\int_0^z \frac{k\Tg}{\me c^2} \frac{\sigT \Ne c}{H (1+z)} \id z \approx \pot{4.9}{-11} (1+z)^2,
\end{align}
where $\Ne$ is the free electron number density and $H(z)$ the Hubble parameter. The approximation is valid in the radiation-dominated era, assuming standard cosmological parameters. A similar parameter, $\ye=\int \frac{k\Te}{\me c^2} \sigT\Ne c\id t$, can be defined by replacing the photon temperature with the electron temperature, however, for the standard thermal history the difference only becomes noticeable at late times, $z\lesssim 200$ (see Fig.~\ref{fig:y-parameters}).

The dependence of $y_\gamma$ on redshift is illustrated in Fig.~\ref{fig:y-parameters} for the standard cosmology. The value of $y_\gamma$ determines how efficiently photons are redistributed over energy. Comptonization is very efficient at $z\gtrsim \pot{\rm few}{5}$, with $y_\gamma$ exceeding unity by a large amount. In this regime, a $\mu$-distortion \citep{Sunyaev1970mu} is formed. 

After recombination ($z\lesssim 10^3$), redistribution by electron scattering becomes negligible for photons with CMB energies. However, the amount of Doppler broadening still reaches $\simeq 1\%-10\%$ between $z\simeq 10^3$ and $\simeq 10^4$, an effect that for example is important for the calculation of the cosmological helium recombination lines \citep{Jose2008}. 
At $z\lesssim 10^4$, a well-known $y$-distortion \citep{Zeldovich1969}, characterized by partial energy redistribution of photons,  is formed when the medium is heated. In both aforementioned cases, simple analytic approximation for the photon injection Green's function can be found (see Sect.~\ref{sec:mu_estimate} and Sect.~\ref{sec:y_estimate}). During the $\mu$-$y$ transition era ($10^4\lesssim z \lesssim \pot{3}{5}$), detailed numerical computations are required, as the photon spectrum can be found in partially Comptonized states (Sect.~\ref{sec:hybrid}).

We immediately mention that for very high-energy photons, well above the ionization thresholds of hydrogen and helium, the effective $y$-parameter does not drop abruptly after recombination occurred. This is because for high-energy photons it does not matter if the electrons are bound in atoms or free. Thus, if a high-energy photon interacts with a neutral atom during the post-recombination or reionization era, this leads to extra ionizations and production of non-thermal electrons and subsequent particle cascades for which more detailed computations are required \citep{Chen2004, Slatyer2009, Huetsi2009, Valdes2010}. We shall leave a detailed discussion of this case to future work.

\vspace{-2mm}
\subsubsection{Initial effective photon temperature}
\label{sec:Teq}
At $z=\zin$, the CMB spectrum is assumed to be given by a pure blackbody at temperature $\Tin$. Immediately after the photon injection, the number and energy densities of the photon field are 
\bsub
\label{eq:N_rho_initial}
\beal
\label{eq:N_rho_initial_a}
N_\gamma(z_{\rm i})&=N^{\rm Pl}_\gamma(\Tin)\left[1+\frac{\Delta N_\gamma}{N_\gamma}\right],
\\
\label{eq:N_rho_initial_b}
\rho_\gamma(z_{\rm i})&=\rho^{\rm Pl}_\gamma(\Tin)\left[1+\alpha_\rho\xin\frac{\Delta N_\gamma}{N_\gamma}\right],
\end{align}
\esub
where $\xin=h\nuin/k\Tin$ is the dimensionless injection frequency. The injected number of photons is parametrized by $\Delta N_\gamma/N_\gamma$. The number and energy densities of the CMB blackbody are, respectively, given by $N^{\rm Pl}_\gamma(T)=(8\pi/c^3)(kT/h)^3\,G_2^{\rm Pl}\approx 410\,\cm^{-3}(T/2.726\Kel)^3$ and $\rho^{\rm Pl}_\gamma(T)=(8\pi h/c^3)(kT/h)^4\,G_3^{\rm Pl}\approx 0.26\,\eV\,\cm^{-3}(T/2.726\Kel)^4$, so that $\rho^{\rm Pl}_\gamma\propto T^4$ and $N^{\rm Pl}_\gamma\propto T^3$ and $\rho^{\rm Pl}_\gamma\approx \alpha_\rho^{-1} k T N^{\rm Pl}_\gamma\approx 2.701 k T N^{\rm Pl}_\gamma$.
We assumed injection of photons in a narrow line around $\xin$.
After the injection, the comoving energy density of the photon-baryon system remains constant\footnote{This assumes that there is no other energy release process and that one can neglect the small energy extraction caused by the adiabatic cooling of baryons \citep{Chluba2005, Chluba2011therm} or by extra coupling of dark matter to baryons and photons \citep{Yacine2015DM}.} and due to the large excess in the number of photons over baryons, one can use Eq.~\eqref{eq:N_rho_initial_b} to determine $\Tin$, giving 
\beal
\Tin\approx T_0(1+\zin)\left[1-\frac{\alpha_\rho\xin}{4} \frac{\Delta N_\gamma}{N_\gamma}\right].
\end{align}
Here, we set the effective temperature of the photon field after the photon injection to $\Tg(z)=T_0(1+z)$, where $T_0=(2.726\pm0.001)\,{\rm K}$ is the CMB temperature today \citep{Fixsen1996, Fixsen2009}.
Inserting $\Tin$ into Eq.~\eqref{eq:N_rho_initial_a}, we find 
\beal
\Tin^{N}\approx T_0(1+\zin)\left[1+\left(1-\frac{3\alpha_\rho\xin}{4}\right) \frac{1}{3}\frac{\Delta N_\gamma}{N_\gamma}\right],
\end{align}
which shows that for injection at $\xin\equiv x_0=(4/3)/\alpha_\rho\approx 3.6016$ we obtain $\Tin^{N}\equiv T_0(1+\zin)$. In this case, the added number of photons is exactly the amount required to increase the energy density of the photon field from $\Tin$ to $\Tg(\zin)$ while having both the effective number and energy density of the photon field agree with a vanishing chemical potential. Explicitly this means that a full blackbody distribution can in principle be restored by simply redistributing photons over frequency (smearing out the narrow line). Following \citet{Hu1995PhD}, we call this case {\it balanced photon injection} scenario. As we will see below, a second balanced photon injection regime can be found at very low frequencies, where photons are quickly absorbed and converted into pure heat (Sect.~\ref{sec:mu_estimate}).

\vspace{-2mm}
\subsubsection{Initial Compton equilibrium temperature}
\label{sec:Teq_in}
In our computations, we simply use $\Te=\Tin$ for the initial electron temperature at $\zin$. However, after the photon injection, $\Te$ quickly approaches the Compton equilibrium temperature \citep{Levich1970} in the distorted photon field. Assuming $\delta$-function photon injection, we find
\beal
\Teq&=
\frac{\int h\nu^4 n(1+n)\id \nu}{4k\int \nu^3 n\id \nu}
\approx\Tin\left[1+\int\left(x\,\frac{\expf{x}+1}{\expf{x}-1}-4\right) \frac{x^3 \Delta n(x, z)}{4G_3^{\rm Pl}}\id x\right]
\nonumber\\
&\approx\Tin\left[1+\left(\xin\,\frac{\expf{\xin}+1}{\expf{\xin}-1}-4\right) \frac{\alpha_\rho \xin}{4}\frac{\Delta N_\gamma}{N_\gamma}\right],
\end{align}
with $\alpha_\rho/4\approx 0.0925$, $x=h\nu/k\Tin$ and the distortion of the photon occupation number, $\Delta n$, which for single photon injection initially reads $\Delta n=G_2^{\rm Pl} x^{-2}\delta(x-\xin)\,\Delta N_\gamma/N_\gamma$. This expression shows that for $\xin\ll 1$ we have $\Teq \approx \Tin[1-(\alpha_\rho \xin/2)\Delta N_\gamma/N_\gamma]<\Tin$, while for $\xin\gg 1$ we obtain $\Teq \approx \Tin[1+(\alpha_\rho \xin^2/4)\Delta N_\gamma/N_\gamma]>\Tin$. For $\xin\approx 3.83$, we find $\Teq\approx \Tin$, which at low redshifts ($z\lesssim \pot{5}{4}$) defines the transition regime between net heating ($\xin>3.83$) and cooling ($\xin<3.83$) caused by photon injection (Sect.~\ref{sec:y_estimate}). 

Although by adding photons to the spectrum we increase the effective temperature of the photon field, energy exchange with the electrons does not necessarily lead to net heating of the plasma. Photons injected at low frequencies up-scatter and thus cool the electrons ($+$baryons), while injection at very high frequencies leads to Compton heating through electron recoil. This energy exchange, respectively, manifests itself in negative and positive $\mu$ and $y$ distortion contributions due to subsequent interactions of the electrons with the CMB blackbody photons. Here, we identify the regimes in redshift and frequency for which net heating and cooling of the thermal plasma occurs (see Fig.~\ref{fig:Regimes_II} for summary).

\subsection{Chemical potential era}
\label{sec:mu_estimate}
At high redshifts ($z\gtrsim \pot{3}{5}$), Compton scattering is very efficient in bringing electrons and photons into kinetic equilibrium. In this regime, a pure $\mu$-distortion \citep{Sunyaev1970mu, Illarionov1975} is formed. Neglecting the effect of DC scattering and BR, the chemical potential and final CMB temperature can be determined from Eq.~\eqref{eq:N_rho_initial} using the conditions $N^{\rm BE}_\gamma(z_{\rm f})\approx N_\gamma(z_{\rm i})$ and $\rho^{\rm BE}_\gamma(z_{\rm f})\approx \rho_\gamma(z_{\rm i})$ for a Bose-Einstein spectrum\footnote{These expressions are found by integrating the Bose-Einstein occupation number, $n_{\rm BE}=1/[\expf{x/(1+\Delta T/T_{\rm i})+\mu_0}-1]$, and expanding in $\Delta T/T_{\rm i}$ and $\mu_0$.} 
\bsub
\label{eq:N_rho_final}
\beal
\label{eq:N_rho_final_a}
N^{\rm BE}_\gamma(z_{\rm f})
&
\approx N^{\rm Pl}_\gamma(T_{\rm i})\left[1+3\frac{\Delta T}{T_{\rm i}}-\mathcal{M}^{\rm c}_2\,\mu_0\right],
\\
\label{eq:N_rho_final_b}
\rho^{\rm BE}_\gamma(z_{\rm f})
&
\approx\rho^{\rm Pl}_\gamma(T_{\rm i})\left[1+4\frac{\Delta T}{T_{\rm i}}-\mathcal{M}^{\rm c}_3\,\mu_0\right],
\end{align}
\esub
where $\mathcal{M}^{\rm c}_2\approx 1.3684$ and $\mathcal{M}^{\rm c}_3\approx 1.1106$. 
Introducing the constant, $\kappa^{\rm c}=4\mathcal{M}^{\rm c}_2-3\mathcal{M}^{\rm c}_3\approx 2.1419$, the solution of this system thus is
\bsub
\label{eq:mu_sol}
\beal
\label{eq:mu_sol_a}
\mu_0(\zin)
&
\approx \frac{3}{\kappa^{\rm c}}
\left[\frac{\Delta \rho_\gamma}{\rho_\gamma}-\frac{4}{3}\frac{\Delta N_\gamma}{N_\gamma}\right]
= 
\frac{3\alpha_\rho}{\kappa^{\rm c}}\Big[x_{\rm i}-x_0\Big]\frac{\Delta N_\gamma}{N_\gamma}
\nonumber\\
&\approx 0.5185  \Big[x_{\rm i}-3.6016\Big] \frac{\Delta N_\gamma}{N_\gamma},
\\[1mm]
\frac{\Delta T(\zin)}{T_{\rm i}}
&\approx \frac{\mathcal{M}^{\rm c}_2}{\kappa^{\rm c}}\frac{\Delta \rho_\gamma}{\rho_\gamma}
-\frac{\mathcal{M}^{\rm c}_3}{\kappa^{\rm c}}\frac{\Delta N_\gamma}{N_\gamma}
= 
\frac{\mathcal{M}^{\rm c}_2}{3}\mu_0 + \frac{1}{3}\frac{\Delta N_\gamma}{N_\gamma}
\nonumber\\
&\approx 0.4561\mu_0+ \frac{1}{3}\frac{\Delta N_\gamma}{N_\gamma},
\end{align}
\esub
where we used $\Delta \rho_\gamma/\rho_\gamma=\alpha_\rho\,\xin \,\Delta N_\gamma/N_\gamma$ and $x_0=(4/3)/\alpha_\rho$. These expression again show that for photon injection at $\xin< 3.6$ a negative chemical potential is expected \citep{Hu1995PhD}. In this case, photons up-scatter on average, cooling the electrons and baryons and causing energy extraction from the CMB blackbody part to balance the total energy budget. For photon injection at $\xin> 3.6$, the opposite happens and a positive chemical potential is formed. 

We assumed that no extra energy in form of heat was released. Also injecting $\left.\Delta\rho_\gamma/\rho_\gamma\right|_{\rm h}$ into the plasma (energy that heats the electrons or baryons directly without any change in the photon number), we obtain an extra contribution \citep{Sunyaev1970mu}, $\mu_{\rm h}\approx 1.4006 \left.\Delta\rho_\gamma/\rho_\gamma\right|_{\rm h}$, to the resulting chemical potential. Thus, more generally, in Eq.~\eqref{eq:mu_sol} one can interpret $\Delta\rho_\gamma/\rho_\gamma$ as the sum $\Delta\rho_\gamma/\rho_\gamma\equiv \left.\Delta\rho_\gamma/\rho_\gamma\right|_{\rm h}+\alpha_\rho \xin \Delta N_\gamma/N_\gamma$. Then the effective chemical potential and blackbody temperature, formed after a short time, are
\bsub
\label{eq:mu_eff}
\beal
\label{eq:mu_eff_a}
\mu^*_0(\zin)
&
\approx \frac{3}{\kappa^{\rm c}}\!\!\left.\frac{\Delta \rho_\gamma}{\rho_\gamma}\right|_{\rm h}
+\frac{3\alpha_\rho}{\kappa^{\rm c}}\Big[x_{\rm i}-x_0\Big]
\frac{\Delta N_\gamma}{N_\gamma}
\\[1mm]
\frac{\Delta T^\ast(\zin)}{T_{\rm i}}
&\approx \frac{\mathcal{M}^{\rm c}_2}{3}\mu^\ast_0 + \frac{1}{3}\frac{\Delta N_\gamma}{N_\gamma}.
\end{align}
\esub
This shows that due to the extra heating the total effective chemical potential can in principle vanish or be positive, even if photon injection causes a net negative value for $\mu$. This also implies that for both photon injection and energy release, an interpretation of the final distortion constraints becomes more ambiguous.

\vspace{1mm}
\subsubsection{Including photon emission and absorption}
\label{sec:mu_em_abs}
After the initial evolution of the photon distribution and electron temperature reaches a quasi-stationary state, the amplitude of $\mu$ slowly reduces due to DC and BR emission. The final, observable chemical potential today is then expected to be roughly given by 
\beal
\label{eq:mu_obs}
\mu_0(z=0) &\approx \mu^*_0(\zin) \, \Jbb(\zin),
\end{align}
where $\Jbb(\zin)$ is the {\it distortion visibility function} \citep{Chluba2011therm}, which can be approximated as $\Jbb(\zin)\approx \expf{-(\zin/\zmu)^{5/2}}$, with thermalization redshift $\zmu\approx \pot{1.98}{6}$ \citep[e.g.,][]{Burigana1991, Hu1993}. This factor takes the subsequent thermalization process for the $\mu$-distortion into account, leading to a strong suppression of the resulting chemical potential caused by photon injection or energy release at $\zin\gg \zmu$. More accurate expressions for the visibility function can be given \citep{Khatri2012b, Chluba2014}. Using the results of \cite{Chluba2014}, we find 
\beal
\mathcal{J}^\ast_{\rm bb}(\zin)\approx 0.983\,\expf{-(z/\zmu)^{2.5}}\Big[1 - 0.0381(z/\zmu)^{2.29}\Big]
\end{align}
to represent the distortion visibility at $\pot{3}{5}\lesssim \zin \lesssim\pot{6}{6}$ very well for the standard cosmology. We neglected relativistic temperature corrections due to Compton scattering, which become noticeable at $\zin\gtrsim \pot{4}{6}$ \citep{Chluba2014}.

\vspace{1mm}
\subsubsection{Improvement at low frequencies}
\label{sec:mu_em_abs_II}
While the above analysis captures the basic features for photon injection in the $\mu$-era, the situation is slightly more complicated at very low frequencies ($x\lesssim 10^{-4}-10^{-3}$). The most important difference is that for photon injection at very low frequencies, DC and BR are very efficient and thus absorb extra photons before they can be transported towards higher frequencies by Compton scattering. In this case, photon injection again becomes similar to pure energy release, resulting in a small positive chemical potential, although from Eq.~\eqref{eq:mu_sol} and \eqref{eq:mu_obs} a negative chemical potential $\mu_0\approx - 1.87 \mathcal{J}^\ast_{\rm bb}(\zin)\,\Delta N_\gamma/N_\gamma$ is expected for $\xin\ll 3.6$. 

\begin{figure}
\centering
\includegraphics[width=1.03\columnwidth]{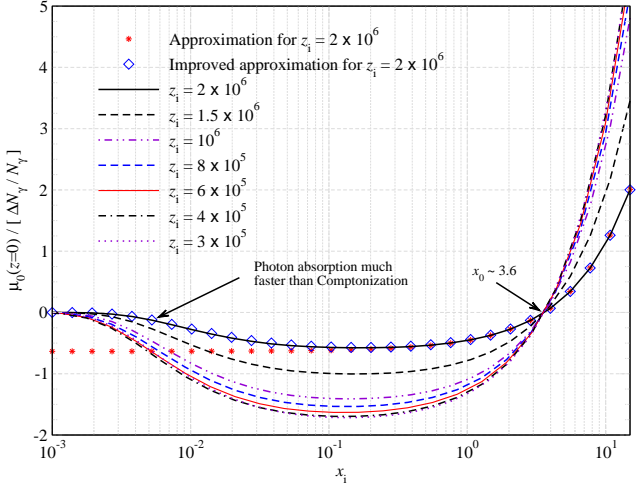}
\caption{Final chemical potential, $\mu_0(z=0)$, after single photon injection at $\xin$ for different $\zin$. For comparison, we show the simple approximation $\mu_0(z=0)\approx (3\alpha_\rho/\kappa^{\rm c})\left[\xin-x_0\right]\, \mathcal{J}^\ast_{\rm bb}(\zin)\,\Delta N_\gamma/N_\gamma$, with distortion visibility function $\mathcal{J}^\ast_{\rm bb}(\zin)\approx 0.983\,\expf{-(z/\zmu)^{2.5}}\left[1 - 0.0381(z/\zmu)^{2.29}\right]$, as well as the improved approximation, Eq.~\eqref{eq:mu_eff_mod}.}
\label{fig:mu_in_final}
\end{figure}
We illustrate this aspect in Fig.~\ref{fig:mu_in_final}, where we present the numerical results for the final chemical potential obtained with {\tt CosmoTherm} after photon injection at different frequencies and redshifts. As expected, for $\xin\simeq 3.6$, the net distortion vanishes, while $\mu>0$ at $\xin>3.6$. At $\xin<3.6$, the chemical potential is negative, however, instead of approaching $\mu_0\approx - 1.87 \mathcal{J}^\ast_{\rm bb}(\zin)\,\Delta N_\gamma/N_\gamma$ at $\xin\ll 3.6$, $\mu$ returns to zero and again becomes positive\footnote{We confirmed this statement although it is not directly visible in Fig.~\ref{fig:mu_in_final}.}.

To understand this aspect in more detail, let us consider the evolution of the injected photons in two steps. Compton scattering broadens the initial narrow line due to the Doppler effect. The average photon energy is furthermore affected by Doppler boosting, electron recoil and simulated recoil, all processes that can be described using the Kompaneets equation \citep{Kompa56}. Since the number of photons does not change through Compton scattering, the expressions given above are directly applicable.

Once we include photon emission and absorption terms, a part of the injected photons can disappear before reaching the quasi-stationary $\mu$-evolution phase. Through the absorption process, their energy is immediately converted into heat and thus becomes equivalent to pure energy release. This effect can be incorporated by defining the {\it photon survival probably}, $\Jsurv(\xin, \zin)$, which determines the fraction of injected photons, $\left.\Delta N_\gamma/N_\gamma\right|_{\rm s}=\Jsurv(\xin, \zin)\Delta N_\gamma/N_\gamma$, that {\it survive} the evolution towards the initial $\mu$-phase. Here, it is assumed that this transition occurs rather fast before the main thermalization of the $\mu$-distortion occurs. It is furthermore assumed that the DC emission caused by the injected photons is negligible, an approximation that we find to be valid in the discussion below (see Sect.~\ref{sec:gamma_injection}).
The number of injected photons that were absorbed and converted into heat then is $\left.\Delta N_\gamma/N_\gamma\right|_{\rm d}=\left[1-\Jsurv(\xin, \zin)\right]\Delta N_\gamma/N_\gamma$. Since photon emission and absorption are most important at low frequencies, it is clear that $\Jsurv(\xin, \zin)\rightarrow 0$ for $\xin\rightarrow 0$. On the other hand, photons injected at high frequencies have a high survival probability, so that  $\Jsurv(\xin, \zin)\rightarrow 1$ for $\xin\rightarrow \infty$. We will discuss the precise shape for $\Jsurv(\xin, \zin)$ below (Sect.~\ref{sec:J_x_z_vis}).

Even if the specific moment of absorption after the injection event determines how much energy the absorbed photon has, until that very moment a corresponding energy was already extracted from the thermal background (i.e., the bulk of CMB blackbody photons). Thus, the net energy injected by the absorption event, assuming that overall the evolution towards a $\mu$-distortion occurs very rapidly, simply is $\left.\Delta \rho_\gamma/\rho_\gamma\right|_{\rm d}\approx \alpha_\rho \xin \left[1-\Jsurv(\xin, \zin)\right]\Delta N_\gamma/N_\gamma$. Hence, the total effective energy density that has to be thermalized is still $\Delta\rho_\gamma/\rho_\gamma\equiv \left.\Delta\rho_\gamma/\rho_\gamma\right|_{\rm d}+\left.\Delta\rho_\gamma/\rho_\gamma\right|_{\rm s}=\alpha_\rho \xin\,\Delta N_\gamma/N_\gamma$, where $\left.\Delta\rho_\gamma/\rho_\gamma\right|_{\rm s}=\alpha_\rho \xin\,\Jsurv(\xin, \zin)\Delta N_\gamma/N_\gamma$. However, the effective number of extra photons that need to be ingested is reduced to 
\beal
\label{eq:N_eff}
\frac{\Delta N^*_\gamma}{N_\gamma}
\approx \left.\frac{\Delta N_\gamma}{N_\gamma}\right|_{\rm s}
\approx  \Jsurv(\xin, \zin) \frac{\Delta N_\gamma}{N_\gamma}.
\end{align}
This means that Eq.~\eqref{eq:mu_eff_a} is modified to
\beal
\label{eq:mu_eff_mod}
\mu^*_0(\zin)
&
\approx \frac{3}{\kappa^{\rm c}}\!\!\left.\frac{\Delta \rho_\gamma}{\rho_\gamma}\right|_{\rm h}
+\frac{3\alpha_\rho}{\kappa^{\rm c}}\Big[x_{\rm i}-x_0\, \Jsurv(\xin, \zin)\Big]
\frac{\Delta N_\gamma}{N_\gamma},
\end{align}
which shows that for $\xin\ll 1$ (or specifically $x_0\, \Jsurv(\xin, \zin)< \xin$) photon injection again leads to a positive chemical potential contribution, while at high frequencies the expression in Eq.~\eqref{eq:mu_eff_a} is recovered. Using the approximation Eq.~\eqref{eq:P_QS} for $\Jsurv(\xin, \zin)$, we find very good agreement with our numerical result for $\mu^*_0$ (see Fig.~\ref{fig:mu_in_final}). The approximation slightly degrades towards lower redshifts since the transition to the quasi-stationary $\mu$-phase no longer is quasi-instantaneous, however, for estimates this expression suffices. More detailed estimates can directly rely on the numerical results obtained for different injection redshifts and frequencies.

\subsubsection{Photon injection Green's function in the $\mu$-era}
Putting all this together, we have the simple approximation of the photon injection Green's function in the $\mu$-era
\beal
\label{eq:G_x_z_muera}
G_{\rm in}(\nu, \nu', z)
&\!\approx \!\frac{3\alpha_\rho}{\kappa^{\rm c}}\Big[x'-x_0\, \Jsurv(x', z) \Big]\, \Jbb^*(z)\, M(\nu) 
+  \frac{\lambda}{4}\,G(\nu)
\end{align}
where $M(\nu)\approx G(\nu)\left[0.4561 - 1/x\right]$ is the spectrum of a $\mu$-distortion with $G(\nu)=(2h\nu^3/c^2)\,x\expf{x}/(\expf{x}-1)^2$ and $x=h\nu/kT_0$. We added a temperature shift term, $\propto G(\nu)$, since it is clear that the part of the initial energy that is not stored by the $\mu$-distortion has been fully thermalized. We can determine $\lambda(\nu',z)$ using the normalization condition, Eq.~\eqref{eq:G_norm}, yielding
\beal
\label{eq:lambda_x_z_muera}
\lambda(\nu',z)
&\!\approx \alpha_\rho \left(x'- \Big[x' -x_0\, \Jsurv(x', z) \Big]\,\Jbb^*(z)\right).
\end{align}
It is straightforward to confirm that for the single injection source function, $S(\nu, z)=h\nu\, \Delta N_\gamma(z)/N_\gamma(z) \,\delta(\nu-\nu_{\rm i})\delta(z-\zin)$, with $(4\pi/c)\int M(\nu)\id\nu=(\kappa^{\rm c}/3)\,\rho_\gamma(T_0)$ and $(4\pi/c)\int G(\nu)\id\nu=4\,\rho_\gamma(T_0)$ and Eq.~\eqref{eq:Greens_injection} one obtains the correct chemical potential and temperature shift for the total spectrum.

\subsubsection{Photon survival probability in the $\mu$-era}
\label{sec:J_x_z_vis}
How can we determine the photon survival probability, which turned out to be so useful for the description of the distortion caused by photon injection? Clearly, this probability is determined by the competition between Compton scattering and DC and BR emission. For this purpose, we can look at a simpler photon evolution problem, including only Compton scattering (at $\Te=\Tg$) and absorption. The evolution of the distortion of the photon occupation number, $\Delta n(x, y)$, is then given by \citep[e.g.,][]{Chluba2011therm}
\beal
\label{eq:Dn_equation}
\frac{\partial \Delta n}{\partial y}
\approx \frac{1}{x^2}\frac{\partial}{\partial x} x^4 \left[\frac{\partial}{\partial x} \Delta n +\Delta n(1+2 n_{\rm bb}) \right]
- \frac{\Lambda(x, y)}{x^3 n_{\rm bb} \Thg} \Delta n,
\end{align}
where $ n_{\rm bb}=1/(\expf{x}-1)$, $\Thg=k\Tg/\me c^2$ and $\Lambda(x, y)$ determines the DC and BR emission rate.  From Eq.~\eqref{eq:y_gamma}, $y(\zin, z)=y_\gamma(\zin)-y_\gamma(z)$ between two redshifts $\zin$ and $z<\zin$.

\begin{figure}
\centering
\includegraphics[width=\columnwidth]{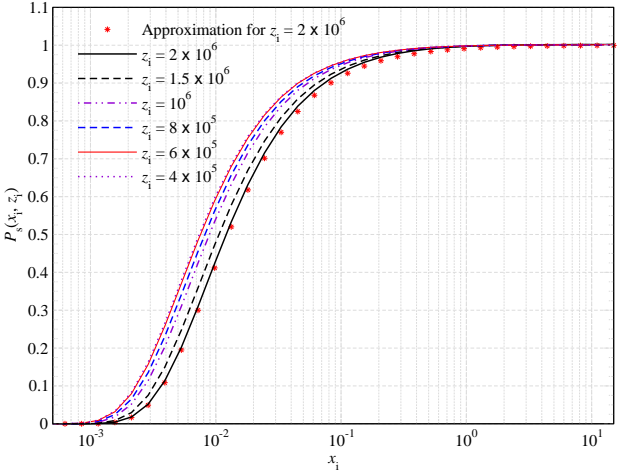}
\caption{Survival probability for different injection frequencies and redshifts in the $\mu$-era. The curves were computed using {\tt CosmoTherm}. The simple approximation, $\Jsurv(x, z)\approx \expf{-\xc(z)/x}$ is also shown for $\zin=\pot{2}{6}$.}
\label{fig:P_surv}
\end{figure}

For photon injection at high frequencies ($x\gtrsim 1$), the absorption term can be neglected until the photon distribution is sufficiently smeared out over the whole frequency range through Compton scattering. Once this regime is reached, the standard thermalization process starts, so that the survival probability until this phase is very close to unity (Fig.~\ref{fig:P_surv}). 

In the other extreme ($x \ll 1$ and $y\lesssim 1$), photons are so rapidly absorbed that Comptonization has not even started to become important before most photons are already gone. In this regime, the survival probability is given by 
\bsub
\label{eq:P_low}
\beal
\Jsurv(x, z)&\approx \expf{-\tau_{\rm abs}(x, z)}
\\
\tau_{\rm abs}(x, z)&\approx \int_0^z\,\frac{\Lambda(x, z')}{x^2} \frac{\sigT \Ne c}{H (1+z')} \id z'.
\end{align}
\esub
Assuming that BR is negligible, one has $\Lambda(x, z)\approx (4\alpha/3\pi)\,\Thg^2 \mathcal{I}_{\rm dc}$, where $\mathcal{I}_{\rm dc}=\int x^4 n_{\rm bb}(1+n_{\rm bb})\id x\approx 25.976$ determines the DC emissivity of the CMB blackbody field \citep{Lightman1981, Thorne1981} and $\alpha\approx 1/137$ is the fine-structure constant. With this, we find
\beal
\label{eq:tau_dc}
\tau_{\rm abs, dc}(x, z)&\approx \frac{\pot{1.2}{-21}}{x^2} (1+z)^3 \approx \frac{\pot{3.5}{-6} }{x^2}y_\gamma^{3/2}.
\end{align}
Strictly speaking, this expression is only valid at low redshift ($z\lesssim \pot{5}{4}$), when Comptonization is inefficient and BR absorption dominates (contradicting our assumptions), an aspect we will return to below. In particular, it strongly overestimates the absorption optical depth during the $\mu$-era, since Compton scattering was completely neglected.

At higher redshifts, Doppler broadening and boosting as well as CMB blackbody-induced stimulated recoil become important at low frequencies. Neglecting photon emission and absorption the evolution of a narrow line. $\Delta n(x, 0)=A\,\delta(x-\xin)/x^2$ is given by \citep{Chluba2008d}
\beal
\label{eq:Dn_sol}
\Delta n(x, y)&=\frac{A}{\sqrt{4\pi y}}\,\frac{\expf{-[\ln(x/\xin)-y]^2/4y}}{x^3}.
\end{align}
The maximum of the distribution thus evolves as $\xin(y)=\xin\,\expf{y}$ \citep{Chluba2008d}, which removes photons from the low-frequency part, where they have a large probability of being absorbed. Because of stimulated scattering, the speed of this motion is smaller than what follows from the classical solution of \cite{Zeldovich1969}, for which one finds $\xin(y)=\xin\,\expf{3y}$. Thus, a simple correction to the absorption optical depth can be obtained using
\beal
\label{eq:tau_abs_corr}
\tau_{\rm abs}(x, z)&\approx \int_0^z\,\frac{\Lambda\left(x\,\expf{y(z,z')}, z'\right)\,\expf{-2y(z,z')}}{x^2} \frac{\sigT \Ne c}{H (1+z')} \id z',
\end{align}
which takes into account that the absorption probability drops as the photon drift towards higher frequencies, but does not account for the effect of Doppler broadening. For DC alone, this gives
\beal
\label{eq:tau_dc_mod}
\tau^*_{\rm abs, dc}(x, z)&\approx 
\frac{\pot{1.9}{-6}}{x^2}\left[\sqrt{2y_\gamma}-F_{\rm D}\left(\sqrt{2y_\gamma}\right)\right],
\end{align}
where $F_{\rm D}(x)=\expf{-x^2}\int^x_0 \expf{y^2}\id y $ is the Dawson integral. For $y_\gamma\ll 1$, one finds $\tau^*_{\rm abs, dc}(x, z)\approx \tau_{\rm abs, dc}(x, z)\propto y_\gamma^{3/2}$, however, for $y_\gamma>1$, one has $\tau^*_{\rm abs, dc}(x, z)\approx (\pot{2.7}{-6}/x^2)\,\sqrt{y_\gamma}$, which increases significantly slower than without the effect of up-scattering. Still, even this improved approximation strongly underestimates the survival probability at low frequencies, since line broadening caused by the Doppler effect, $\Delta \nu/\nu\simeq 2\sqrt{y \ln 2}$, also helps to transport photons towards high frequencies. 
%

It turns out that a simple approximation can be found using the quasi-stationary solution to Eq.~\eqref{eq:Dn_equation}. This yields
\beal
\label{eq:P_QS}
\Jsurv(x, z)&\approx \expf{-\xc(z)/x},
\end{align}
where $\xc(z)=\sqrt{\Lambda(\xc, z)/\Thg}$ is the critical frequency at which Compton scattering takes over the evolution \citep{Burigana1991, Hu1993}. From \cite{Chluba2014}, we have
\bsub
\label{eq:xc_DC_BR_appr}
\beal
\label{eq:xc_DC_appr}
\xc^{\rm DC}
&\approx\sqrt{\frac{4\alpha}{3\pi}\, \Thg \mathcal{I}^{\rm Pl}_4} \approx \pot{8.60}{-3} \left[\frac{1+z}{\pot{2}{6}}\right]^{1/2}
\\
\xc^{\rm BR}&\approx \pot{1.23}{-3}\,\left[\frac{1+z}{\pot{2}{6}}\right]^{-0.672}
\end{align}
\esub
for the critical frequencies of DC and BR, respectively. To percent precision, the total critical frequency is $\xc^2\approx (\xc^{\rm DC})^2+(\xc^{\rm BR})^2$ \citep{Hu1993}, an approximation that is sufficient for our purposes. In Fig.~\ref{fig:P_surv}, we show the comparison of our approximation for injection redshift $\zin=\pot{2}{6}$, finding good agreement with the numerical result. Overall, we find the approximation to be valid at the $\simeq 10\%-20\%$ level for $\pot{3}{5}\lesssim \pot{\rm few}{6}$. 

The solution, Eq.~\eqref{eq:P_QS}, also represents the frequency dependence of the chemical potential, $\mu(z, x)\simeq \mu_0(z)\,\expf{-\xc(z)/x}$ \citep{Sunyaev1970mu} during the quasi-stationary chemical potential evolution phase. This highlights that also there the factor $\expf{-\xc(z)/x}$ should be interpreted as the probability to have a non-zero chemical potential at a given frequency and redshift  in the $\mu$-era through the competition of emission and absorption and scattering.

Equation~\eqref{eq:P_QS} also allows us to estimate the frequency $x_{\rm h}~\ll~1$ at which the chemical potential is expected to become positive again. From the condition $\xin \approx x_0 \Jsurv(x, z)$, we find
\beal
\label{eq:xc_low}
x_{\rm h}(z)&\approx \xc(z)/8.3
\end{align}
to reproduce the numerical results very well (see Fig.~\ref{fig:Regimes_II}). Photon injection around this frequency during the $\mu$-era corresponds to a balanced injection scenario, with vanishing chemical potential. Below that frequency, photons are effectively converted into heat before reaching the high-frequency tail through Comptonization.

\subsection{Compton-$y$ era}
\label{sec:y_estimate}
We now discuss the effect of photon injection on the CMB spectrum during the $y$-distortion era ($z\lesssim 10^4$). At these redshifts, the $y$-parameter is already rather small, $y_\gamma\lesssim 10^{-2}$ (see Fig.~\ref{fig:y-parameters}), so that any smearing of photons over frequency or direct energy exchange with electrons is quite limited. At low frequencies, the photon emission and absorption process is furthermore dominated by BR.

\subsubsection{Photon injection after recombination}
\label{sec:injection_low_rec}
As Fig.~\ref{fig:y-parameters} illustrates, for scenarios with late photon production, after recombination finished ($z\lesssim 10^3$), one can practically neglect Compton scattering, unless the initial photon energy is very large so that electron recoil, $\Delta \nu/\nu\simeq - x y_\gamma$, or ionizations of neutral atoms become significant. Compton scattering remains negligible even if one accounts for the reionization and structure formation processes \citep[e.g.,][]{Hu1994pert, Cen1999, Refregier2000, Oh2003} or heating due to magnetic fields \citep{Jedamzik2000, Sethi2005, Kunze2014, Chluba2015PMF}, the former of which could on average increase the effective $y$-parameter to the level of $\pot{2}{-6}$ from small haloes \citep{Hill2015}. 

Here, we do not consider photon injection at energies above the hydrogen ionization threshold, $x_{\rm H}\simeq \pot{5.8}{4}/(1+z)$, around the Lyman-$\alpha$ excitation frequency, $x_{\rm H, ex}\simeq \pot{4.3}{4}/(1+z)$, or at the corresponding energies of helium. This would affect the ionization history \citep{Peebles2000, Chen2004, Padmanabhan2005} and produce CMB spectral distortions from the reprocessing of photons and heat by atomic species \citep{Chluba2008c, Chluba2010a}. In this case, a more detailed treatment is required \citep{Slatyer2009, Huetsi2009, Valdes2010, Slatyer2015}; these considerations are also relevant when deriving CMB anisotropy constraints on annihilating dark matter particles \citep[e.g.,][]{Galli2009, Huetsi2009, Huetsi2011, Planck2015params}. Similarly, we neglect other interactions with atomic and molecular species at late times (e.g., 21cm absorption, Lyman-Werner bands, etc.).

In this case, the photon distribution only evolves through BR emission and absorption. The kinetic equation for the photon occupation number then reads
\beal
\label{eq:BR_only}
\frac{\partial n}{\partial \tau} 
\approx  \frac{\Lambda_{\rm BR}\,\expf{-\xe}}{\xe^3}\!\left[ 1 - n \, (\expf{\xe}-1)\right] + s(\tau, x),
\end{align}
where $\xe=h\nu/k\Te$ and $\Lambda_{\rm BR}$ determines the BR emissivity. We also included a possible photon source term, $s(\tau, x)$, and defined the Thomson optical depth $\tau=\int \sigT\Ne c\id t$. As this equation shows, even for the CMB alone a distortion arises if the electron temperature differs from the CMB temperature, $\Te\neq\Tg$, when free-free emission tends to bring the CMB spectrum into equilibrium with the electrons at low frequencies, enforcing $n\simeq 1/\xe$ at $\xe\ll 1$. This is independent of any extra photon injection and can be studied separately \citep{Hu1995PhD, Burigana2004ff, Chluba2011therm, Trombetti2014ff}, so that we neglect it below. 

The evolution equation for a distortion to the CMB blackbody then becomes
\beal
\label{eq:BR_only_DF}
\frac{\partial \Delta n}{\partial \tau} 
\approx  - \frac{\Lambda_{\rm BR}(\tau, \xe)(1-\expf{-\xe})}{\xe^3}\,\Delta n+ s(\tau, x).
\end{align}
Between $z=0$ and $\zin$, this equation has the simple solution \citep[compare also,][]{Hu1995PhD}
\bsub
\label{eq:BR_only_DF_sol}
\beal
\label{eq:BR_only_DF_sol_a}
\Delta n (x, z=0)  &\approx  \Delta n(x, \zin)\,\expf{-\tau_{\rm ff}(x, \zin)} + \int_0^{\zin}  \expf{-\tau_{\rm ff}(x, z')} \, \tilde{s}(z', x)  \id z
\\[0mm]
\label{eq:BR_only_DF_sol_b}
\tau_{\rm ff}(x, z)&=\int^z_0 \frac{\Lambda_{\rm BR}(z, \xe)(1-\expf{-\xe})}{\xe^3} \,\frac{\sigT \Ne c  \id z}{H (1+z)},
\end{align}
\esub
where we introduced the source function, $\tilde{s}(z, x)$, with respect to redshift. At very low frequencies, BR absorption efficiently destroys photons, converting their energy into heat, which then reappears as a small $y$-distortion at high frequencies. However, since this occurs only at very low frequencies ($x\ll 10^{-4}$), a lot of low-frequency photons need to be injected to have an appreciable effect. 

The solution, Eq.~\eqref{eq:BR_only_DF_sol}, shows that the Green's function in the considered regime can be expressed as
\beal
\label{eq:G_x_z_low}
G_{\rm in}(\nu, \nu', z)
&\approx 
\left[\frac{c\rho_\gamma(T_0)}{4\pi}\,\expf{-\tau_{\rm ff}(x', z)}\delta(\nu-\nu')\right.
\nonumber\\[-0.5mm]
&\qquad\qquad +\left.\left(1-\expf{-\tau_{\rm ff}(x', z)}\right)\frac{Y(\nu)}{4}\right] x' \alpha_\rho,
\end{align}
where $Y(\nu)\approx G(\nu)\left[x\coth(x/2) - 4\right]$ is the $y$-distortion. The last term accounts for the aforementioned small heating of the medium by BR absorption, which usually is negligible. At $z\lesssim 200$, when matter and radiation thermally decouple, the last term is further suppressed. We furthermore mention that for very small, but non-zero, $y$-parameter one can alternatively replace the $\delta$-function using the Compton scattering kernel \citep{Sazonov2000}.

\begin{figure}
\centering 
\includegraphics[width=1.02\columnwidth]{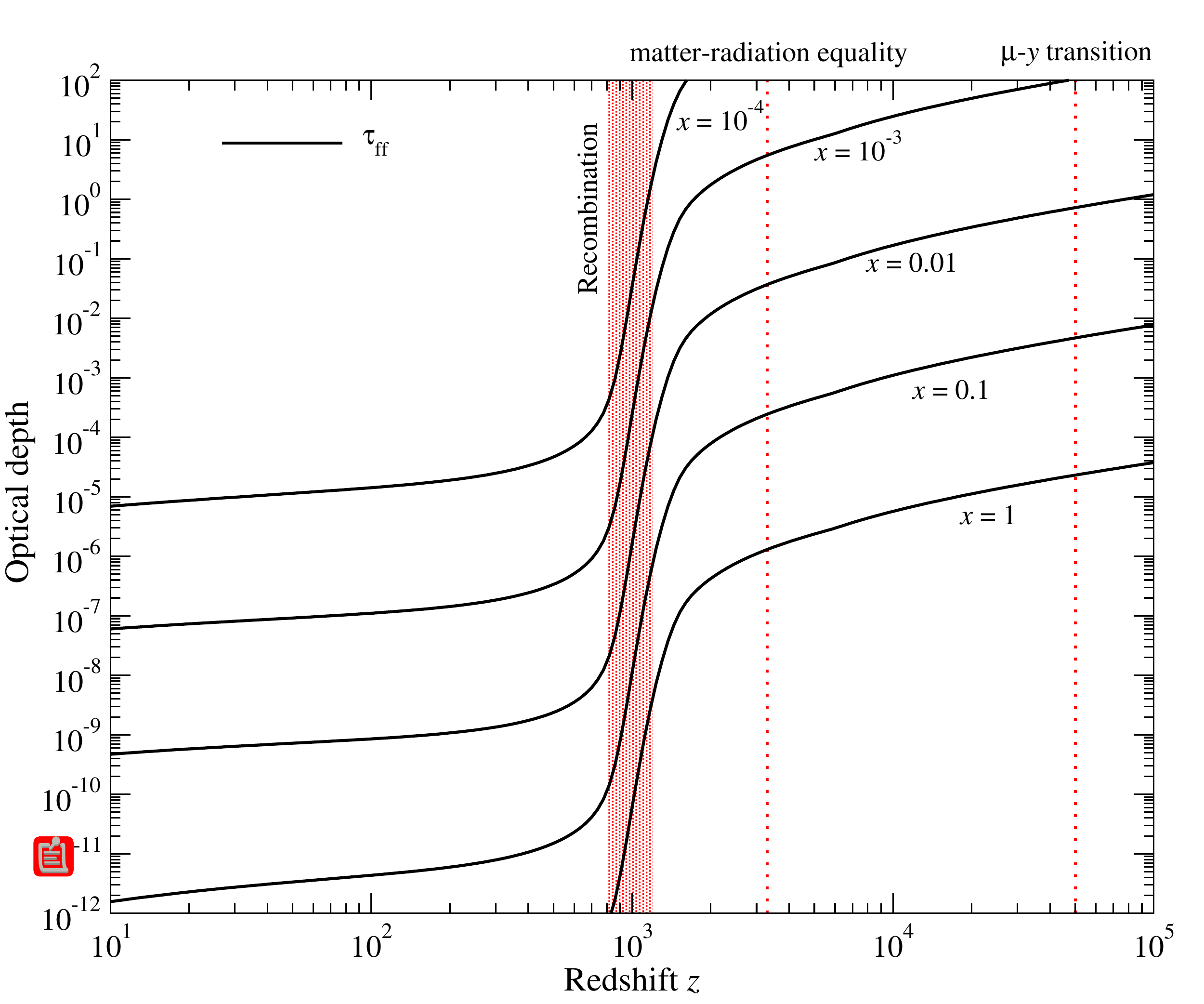}
\caption{Free-free absorption optical depth at different redshifts and frequencies $x$. At low frequencies, the scaling is $\tau_{\rm ff}\simeq F(z)\ln(2.25/x)/x^{2}$. For $x\simeq 10^{-4}\,(\equiv 6 \,{\rm MHz})$, the Universe becomes transparent ($\tau_{\rm ff}\simeq 1$) around recombination. For $x\simeq 10^{-3}$ and $x\simeq 0.01$, this transition happens at $z\simeq 1700$ and $\simeq 10^5$, respectively.}
\label{fig:tau_ff}
\end{figure}

\begin{figure*}
\centering
\includegraphics[width=1.04\columnwidth]{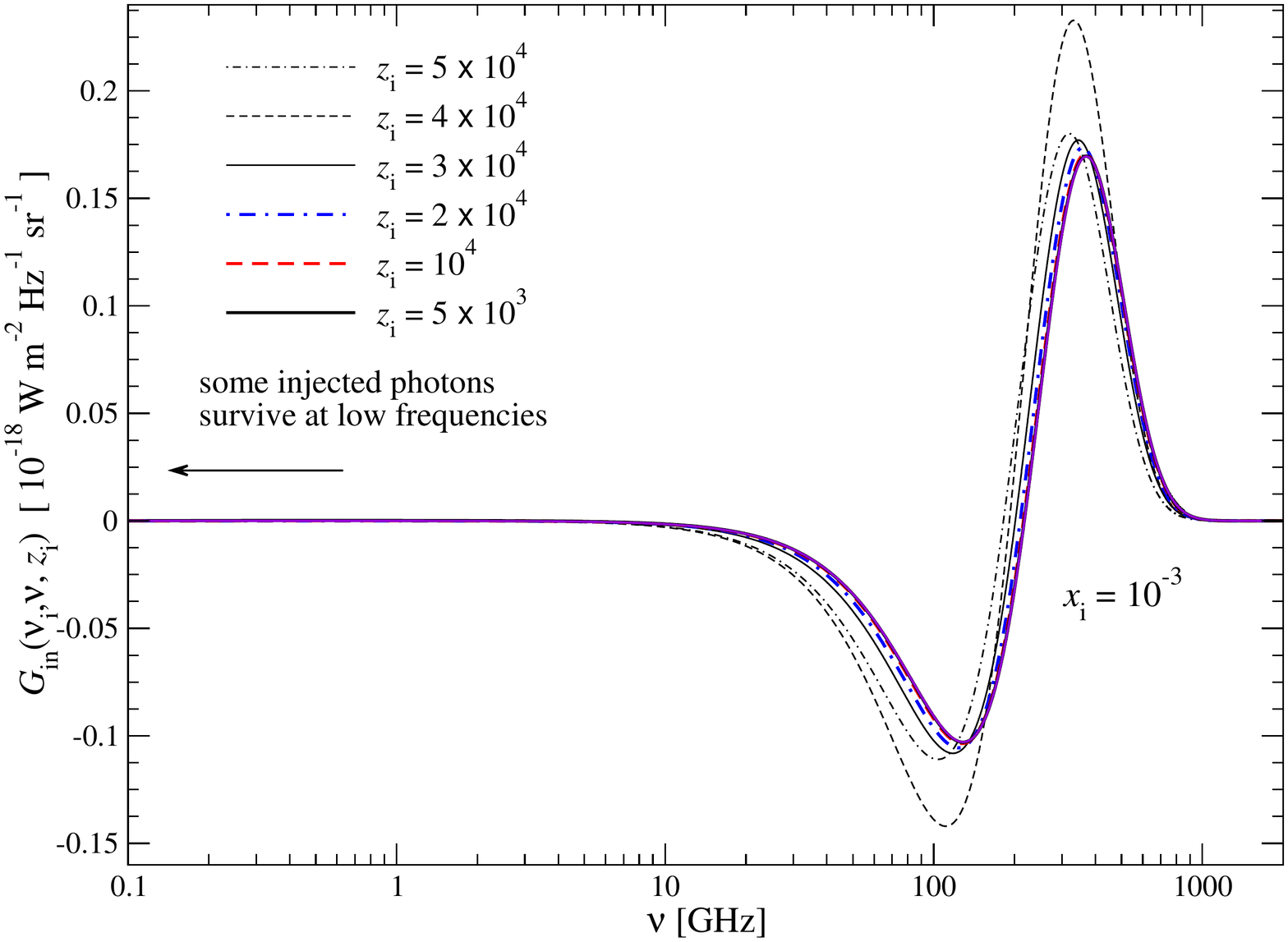}
\includegraphics[width=1.04\columnwidth]{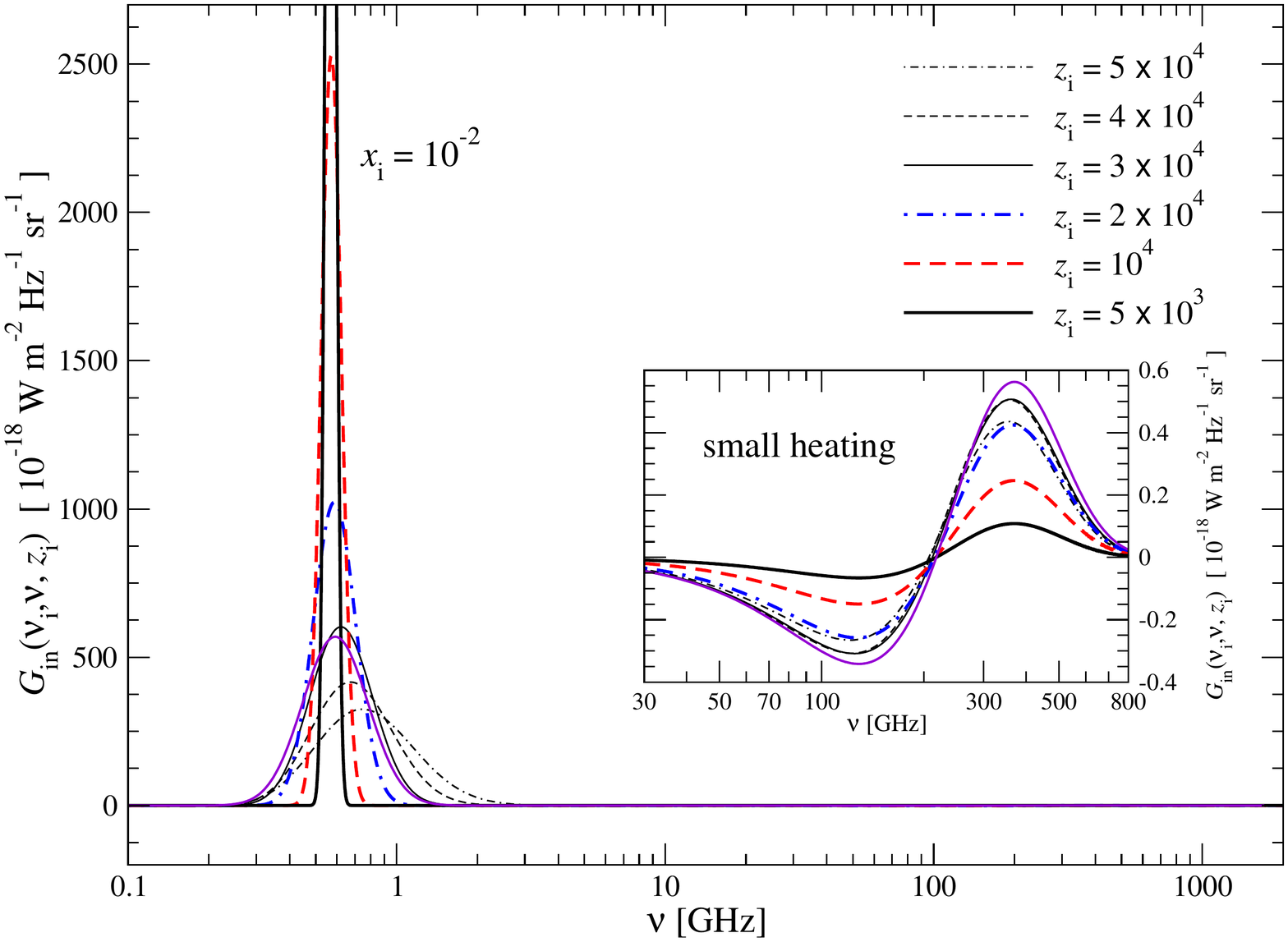}
\\[2mm]
\includegraphics[width=1.04\columnwidth]{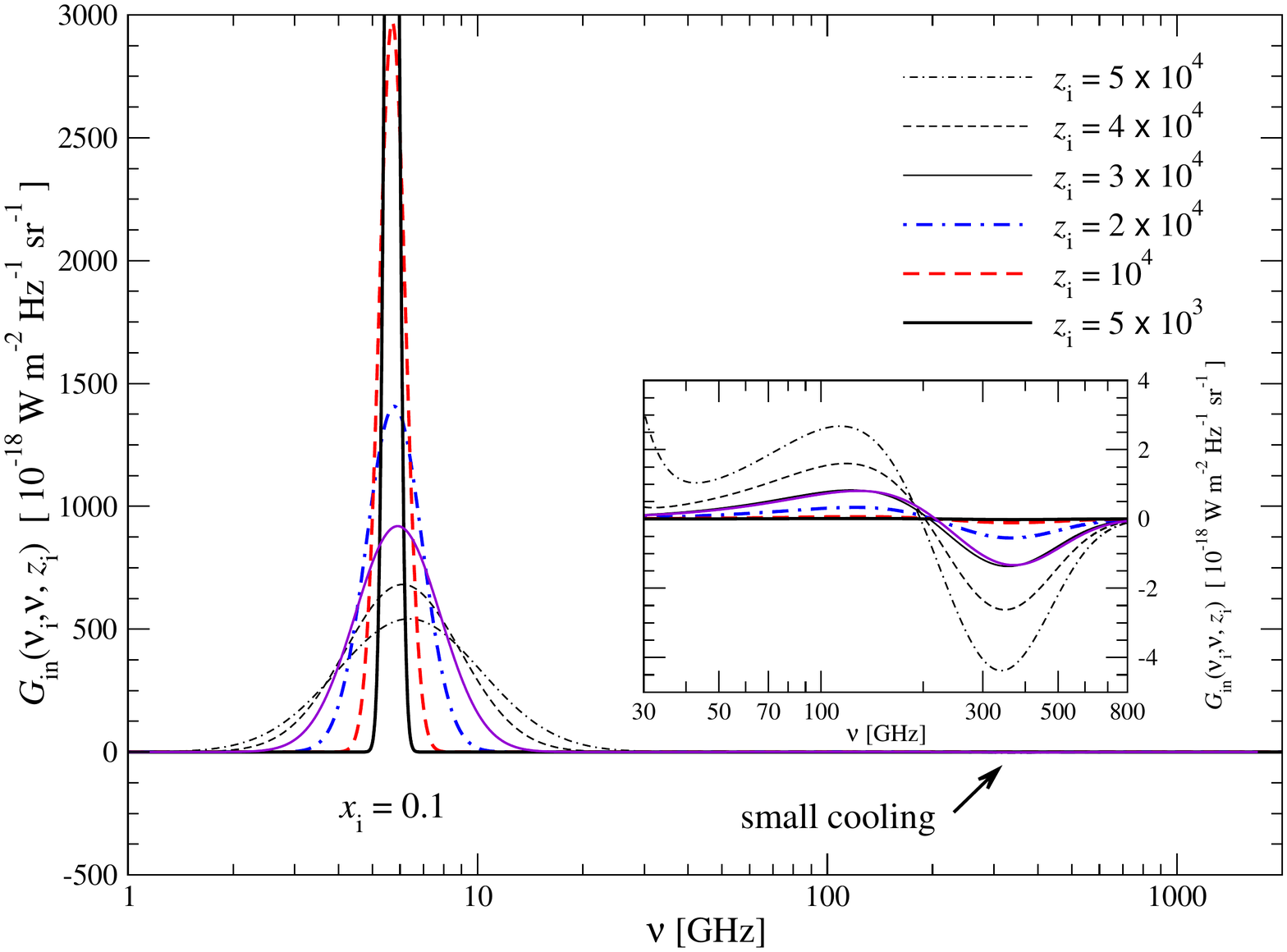}
\includegraphics[width=1.04\columnwidth]{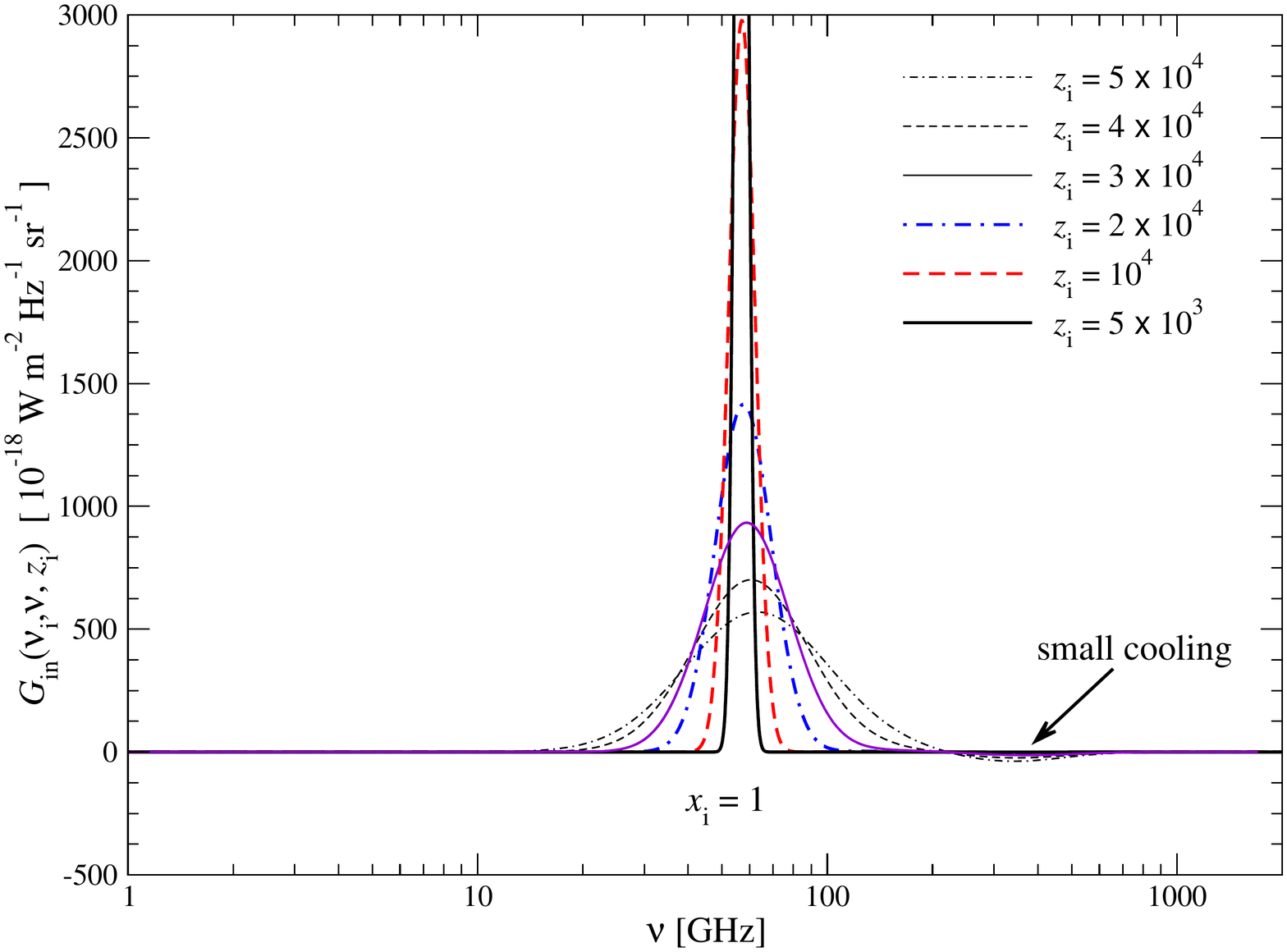}
\\[2mm]
\includegraphics[width=1.04\columnwidth]{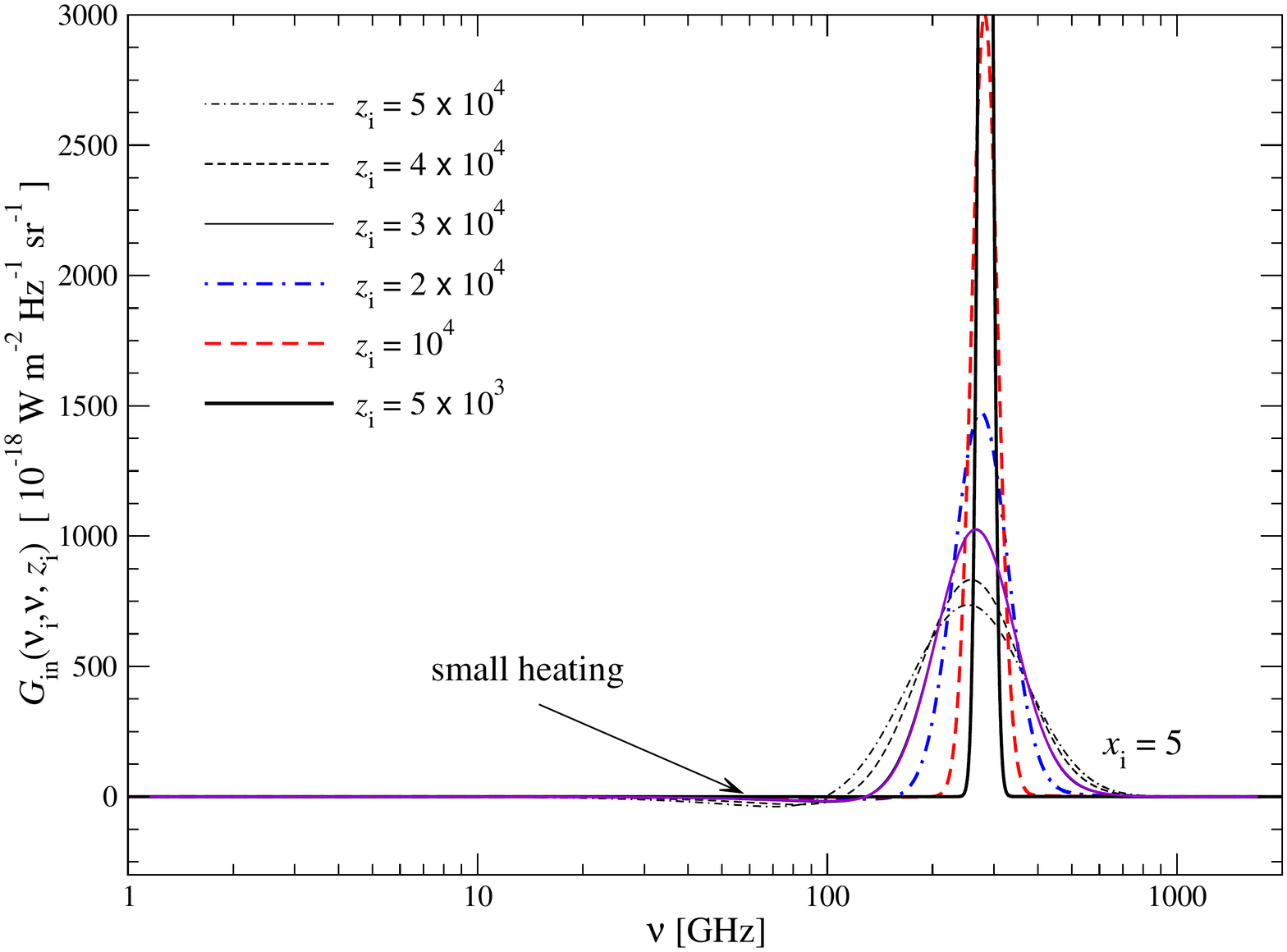}
\includegraphics[width=1.04\columnwidth]{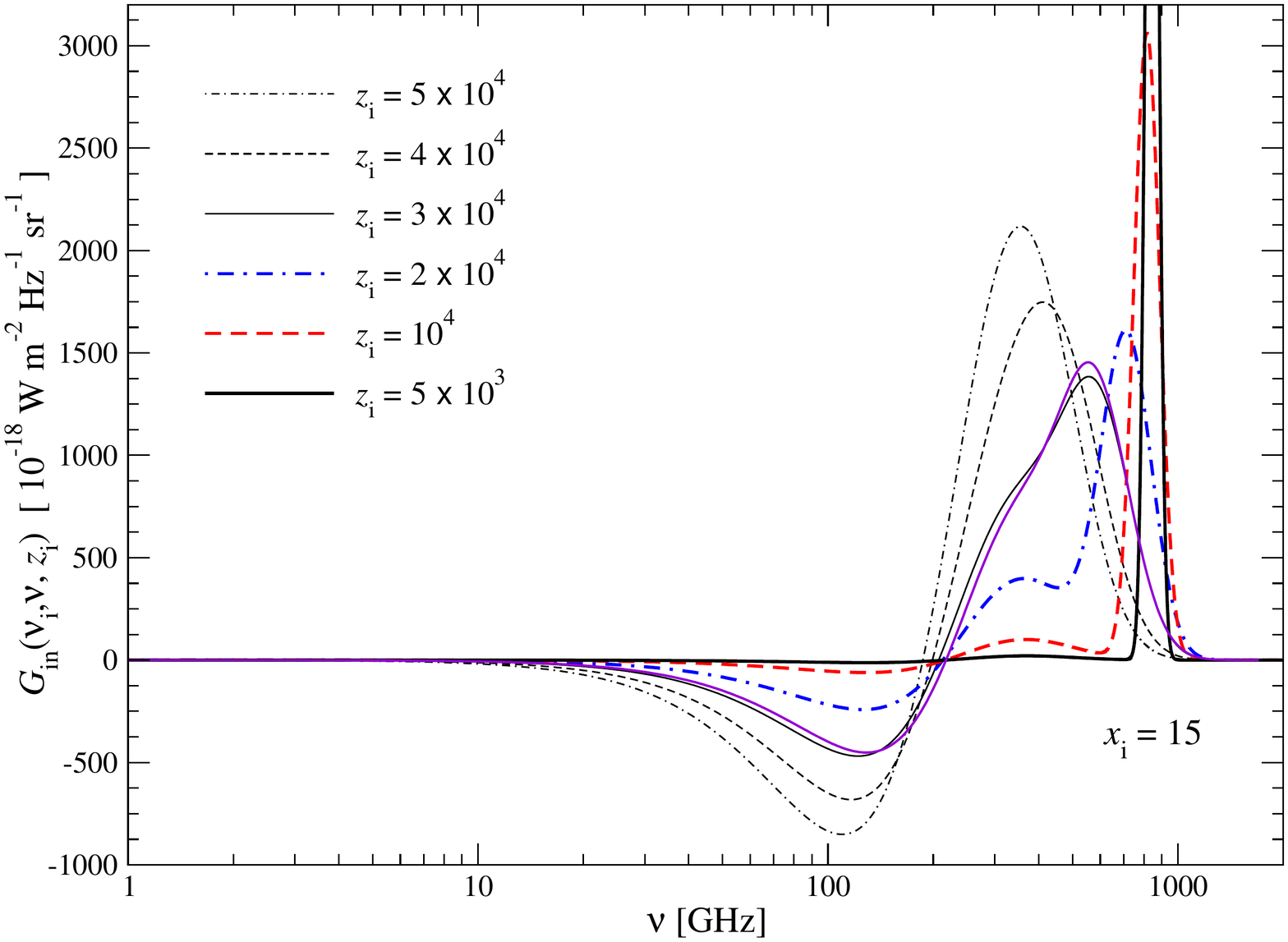}
\caption{Photon injection Green's function for injection at low redshifts, $\zin\lesssim \pot{5}{4}$. In purple we show the analytic approximations for the Green's function using Eq.~\eqref{eq:G_x_z_1e3_1e4} for $x\lesssim 1$ and Eq.~\eqref{eq:G_x_z_1e3_1e4_high} for higher frequencies. Notice that for several cases the approximation fully covers the numerical result. Usually, the high-frequency $y$-type distortion contribution is relatively small, unless photons are injected at $\xin\gtrsim1/y_\gamma$, for which heating through recoil becomes important (e.g., see panel for $\xin=15$). To convert between $x$ and observing frequency today we may use $x\simeq 0.017 \,(\nu/\GHz)$.}
\label{fig:Sols_low}
\end{figure*}

The free-free absorption optical depth can be calculated using the results from {\tt CosmoRec} \citep{Chluba2010b} for the ionization history and approximations for the free-free Gaunt factors from \cite{Itoh2000}. Alternatively, at low temperatures, one can approximate the free-free Gaunt factor as \citep{Draine2011Book}
\beal
\label{eq:Gaunt_low}
g_{\rm ff}(\xe, \Te)
&\approx 1+\ln\left[5.158+\exp\left(\frac{\sqrt{3}}{\pi}\left[\ln\left(\frac{2.25}{Z_i\xe}\right)+\frac{\ln\The}{2}\right] \right) \right],
\end{align}
where $Z_i$ is the charge of the nucleus. For even simpler estimates, one can use $g_{\rm ff}(\xe)\approx \frac{\sqrt{3}}{\pi}\ln(2.25/\xe)$ at $\xe\lesssim 0.37$ \citep{Hu1995PhD}. 
Neglecting helium, with $\Lambda_{\rm BR}=(\alpha \lambda^3_{\rm e}/2\pi\sqrt{6\pi}) \,N_{\rm p} \,\The^{-7/2}\,g_{\rm ff}(\xe)$, where $\lambda_{\rm e}\approx \pot{2.426}{-10}\,{\rm cm}$ is the electron's Compton wavelength, $N_{\rm p}\approx \Ne$ the free proton number density and $\The=k\Te/\me c^2\approx \Thg$, we finally obtain $\Lambda_{\rm BR}\approx \pot{1.9}{-7}\,X_{\rm e}\,(1+z)^{-1/2}\,\ln(2.25/x)$ at $x\lesssim 0.37$, where $X_{\rm e}$ is the free electron fraction. The free-free optical depth is thus roughly given by, $\tau_{\rm ff}(x, z)\approx F(z) \ln(2.25/x)/x^{2}$, where $F(z)$ is a single redshift-dependent function.

We illustrate the numerical results for $\tau_{\rm ff}(x, z)$ in Fig.~\ref{fig:tau_ff} for several values of $x$. In the post recombination era ($z\lesssim 10^3$), signals produced at $x\gtrsim 10^{-4} \,(\equiv 6 \,{\rm MHz})$ are not significantly attenuated by free-free absorption. For percent-level precision, one does, however, need to include the effect of free-free absorption at $x\lesssim 0.1 \,(\equiv 6\, {\rm GHz})$ at $z\simeq 10^3 - 10^5$, an aspect that, e.g., is important for the low-frequency hydrogen and helium recombination spectrum \citep{Chluba2007}.

\subsubsection{Photon injection at $10^3\lesssim z \lesssim 10^4$ and $x\lesssim1$}
\label{sec:injection_before_rec}
At redshifts $10^3\lesssim z \lesssim 10^4$, the total $y$-parameter can reach the percent level (see Fig.~\ref{fig:y-parameters}). In this case, line broadening through the Doppler effect can be as large as $\simeq 10\%$, but no significant net energy exchange between the injected photon distribution and electrons occurs for $\xin\ll 1/y_\gamma$. Thus, the Green's function for this regime has two parts, one that is sourced by the absorption of photons at low frequencies, where BR is efficient and causes a small $y$-distortion, and the other part related to the slightly scattered and broadened injected photon distribution plus a smaller $y$-distortion due to energy exchange. Both of these aspects can be approximately treated independently.

At low frequencies, BR absorption effectively destroys photons, and the photon survival probability is given by 
\beal
\label{eq:P_BR}
\Jsurv(x, z)&\approx \expf{-\tau_{\rm ff}(x, z)},
\end{align}
with $\tau_{\rm ff}(x, z)\approx F(z) \ln(2.25/x)\,x^{-2}$ from Eq.~\eqref{eq:BR_only_DF_sol_b}. It is straightforward to determine the frequency at which most ($\equiv99\%$) of the injected photon energy is absorbed and converted to a $y$-distortion. At $10^3\lesssim z \lesssim 10^4$, we find this for $x\simeq \pot{\rm few}{-3}$ in agreement with our detailed computations.

At slightly higher frequencies ($0.01\lesssim x\lesssim 1$), we can use the solution Eq.~\eqref{eq:Dn_sol} to account for the effects of electron scattering (Doppler broadening, Doppler boosting and stimulated scatterings). In this regime, the average energy of the photon distribution increases like $\Delta\rho_\gamma(y_\gamma)/\rho_\gamma=(\Delta\rho_\gamma/\rho_\gamma)\,\expf{2y_\gamma}$. The energy required for this increase is extracted from the thermal plasma, which leads to a small negative $y$-distortion with effective $y$-parameter,
\beal
\label{eq:y_up-scatter}
y_{\rm up}(\xin,\zin)&\approx \frac{\alpha_\rho}{4} \xin \left[1-\expf{2y_\gamma(\zin)}\right]\frac{\Delta N_\gamma}{N_\gamma}
\approx-\frac{\alpha_\rho}{2} \xin y_\gamma(\zin) \frac{\Delta N_\gamma}{N_\gamma}.
\end{align}
This counteracts the heating $y$-parameter,
\beal
\label{eq:y_low_x_heating}
y_{\rm h}(\xin,\zin)&\approx \frac{\alpha_\rho}{4} \xin \left[1-\expf{-\tau_{\rm ff}(x, z)}\right]\frac{\Delta N_\gamma}{N_\gamma},
\end{align}
caused by the BR absorption process. To fully include the effect of BR absorption, we simply need to multiply the scattering solution and $y_{\rm up}$ by the survival probability given in Eq.~\eqref{eq:P_BR}. For $x\lesssim 1$, we thus have the Green's function
\beal
\label{eq:G_x_z_1e3_1e4}
G_{\rm in}(\nu, \nu', z)
&\approx 
\left[\frac{c\rho_\gamma(T_0)}{4\pi}\,
\frac{\expf{-\tau_{\rm ff}(x', z)}}{\sqrt{4\pi y_\gamma(z)}\,x'}\,\exp\left(-\frac{[\ln(x/x')-y_\gamma(z)]^2}{4y_\gamma(z)}\right)\right.
\nonumber\\
&\qquad+\left.\left(1-\expf{2y_\gamma(z)}\expf{-\tau_{\rm ff}(x', z)}\right)\frac{Y(\nu)}{4}\right] x' \alpha_\rho.
\end{align}
We find this approximation to work very well as long as corrections to the absorption optical depth caused by Doppler broadening are small (see Fig.~\ref{fig:Sols_low}). In particular, for $\xin \simeq 0.1-1$ the solution works extremely well even until $\zin\simeq \pot{3}{4}$.

The solution in Eq.~\eqref{eq:G_x_z_1e3_1e4} shows that, like in the $\mu$-era, if photons are injected only at very low frequencies, a high-frequency $y$-distortion appears through the net competition of heating (by BR absorption) and cooling (by low-frequency photon up-scattering). While at sufficiently low frequencies BR absorption can extract almost {\it all} the injected photon energy, the cooling caused by scattering is limited to a small fraction $\propto y_\gamma \ll 1$. The transition frequency separating the regions of net heating to net cooling can be estimated with the condition $2y_\gamma(z)\approx \tau_{\rm ff}(x_{\rm h}, z)$, as long as $y_\gamma(z)$ is not too large. For $10^3\lesssim z\lesssim 10^4$, we find $x_{\rm h}\simeq 0.01-0.1$ (see Fig.~\ref{fig:Regimes_II}), in very good agreement with our numerical calculations. At $0.01\lesssim \xin\lesssim 1$, the $y$-type contribution to the distortion caused by energy exchange and absorption remains relatively small.

\subsubsection{Photon injection at $10^3\lesssim z \lesssim 10^4$ and $1<x<30$}
\label{sec:injection_before_rec_high}
To describe the solution at higher frequencies ($1<x<30$), we generally need to resort to numerical solutions. Neglecting recoil, one can use the classical solution \citep{Zeldovich1969}
\beal
\label{eq:Dn_sol_ZS}
\Delta n(x, y)&=\frac{A}{\sqrt{4\pi y}}\,\frac{\exp\left(-[\ln(x/\xin)-3y]^2/4y\right)}{x^3},
\end{align}
which differs from the low-frequency solution, Eq.~\eqref{eq:Dn_sol}, only by the net drift term \citep{Chluba2008d}. The solution for pure recoil (neglecting any line broadening through recoil) simply is
$\Delta n(x, y)=A\,x^{-2}\,\delta[x-\xin(y)]$,
with $\xin(y)=\xin/(1+\xin y)$, which gives a drift $\Delta \nu/\nu\simeq - \xin y$ towards lower frequencies. One simple improved approximation, valid for $\xin y\ll 1$, thus is
\beal
\label{eq:Dn_sol_ZS_improved}
\Delta n^*(x, y)&\!=\!\frac{A}{\sqrt{4\pi y}}\frac{\exp\left(-\left[\ln(x/\xin)-3y+\ln(1+\xin y)\right]^2/4y\right)}{x^3}.
\end{align}
This solution gives $\Delta\rho_\gamma(y_\gamma)/\rho_\gamma=(\Delta\rho_\gamma/\rho_\gamma)\,\expf{4y_\gamma}/(1+\xin y_\gamma)$, which captures the aforementioned effects.

\begin{figure}
\centering
\includegraphics[width=1.07\columnwidth]{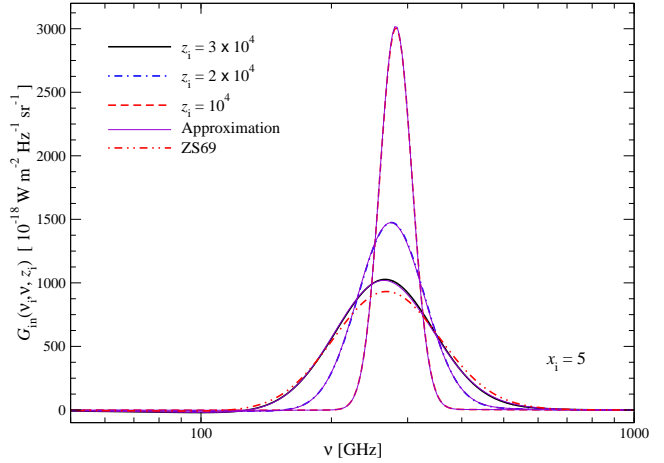}
\caption{Comparison of the approximation in Eq.~\eqref{eq:Dn_sol_ZS_improved_II} with the full numerical results for $\xin=5$ and several injection redshifts. We also show the classical solution, Eq.~\eqref{eq:Dn_sol_ZS}, for $\zin=\pot{3}{4}$, which clearly demonstrates the improvement of the new approximation.}
\label{fig:sol_ill}
\end{figure}

\begin{figure*}
\centering
\includegraphics[width=1.04\columnwidth]{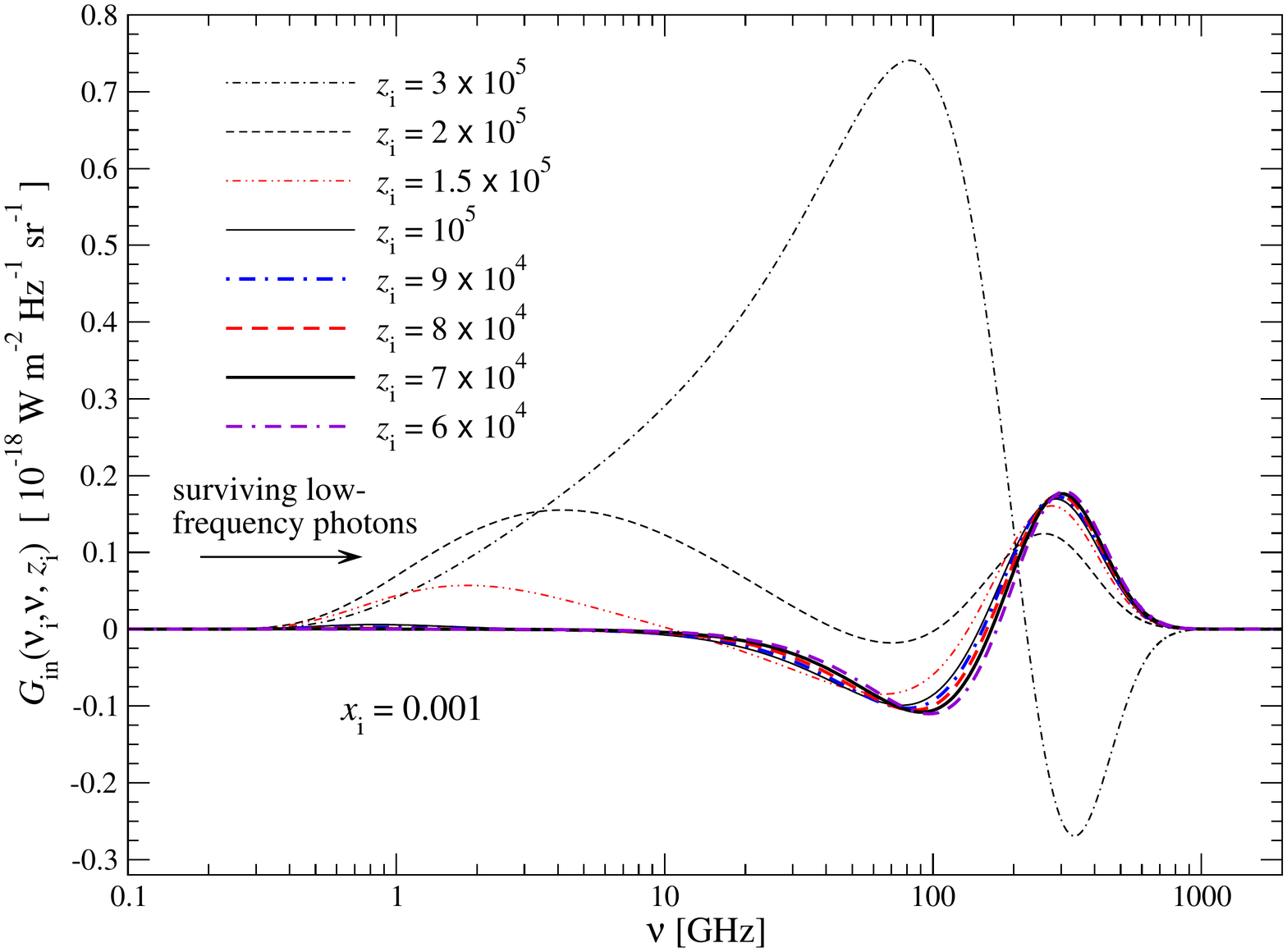}
\includegraphics[width=1.04\columnwidth]{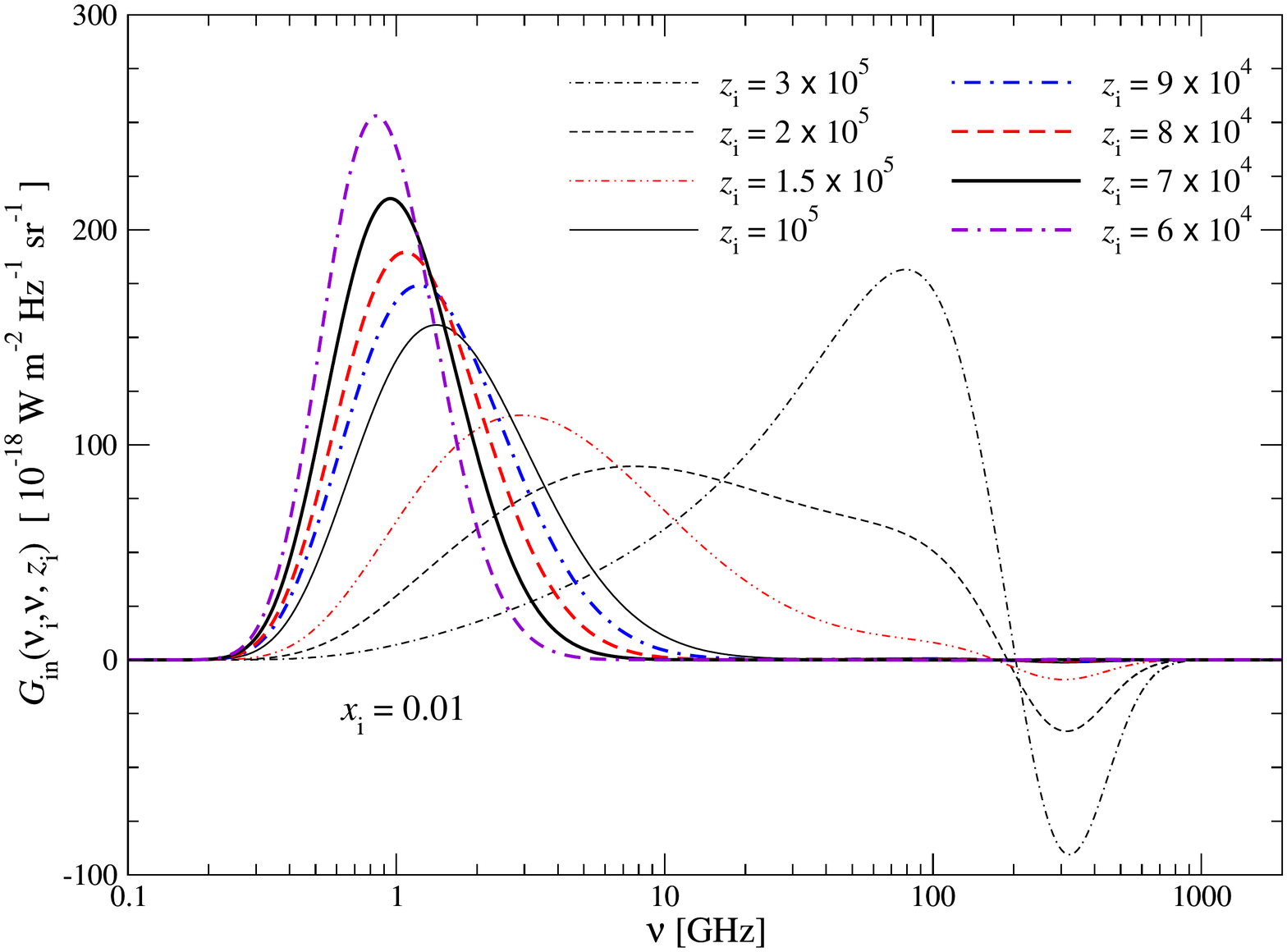}
\\[2mm]
\includegraphics[width=1.04\columnwidth]{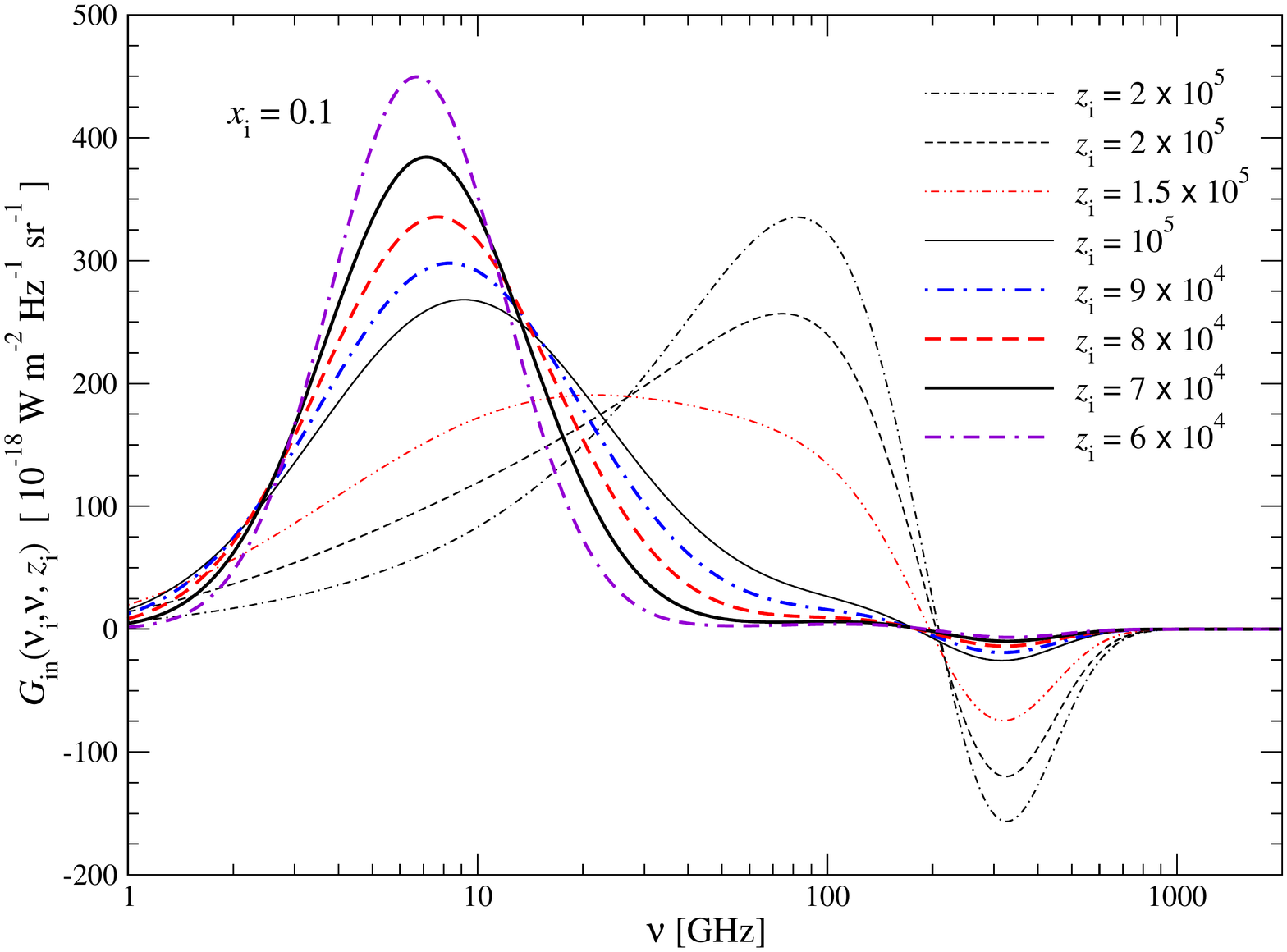}
\includegraphics[width=1.04\columnwidth]{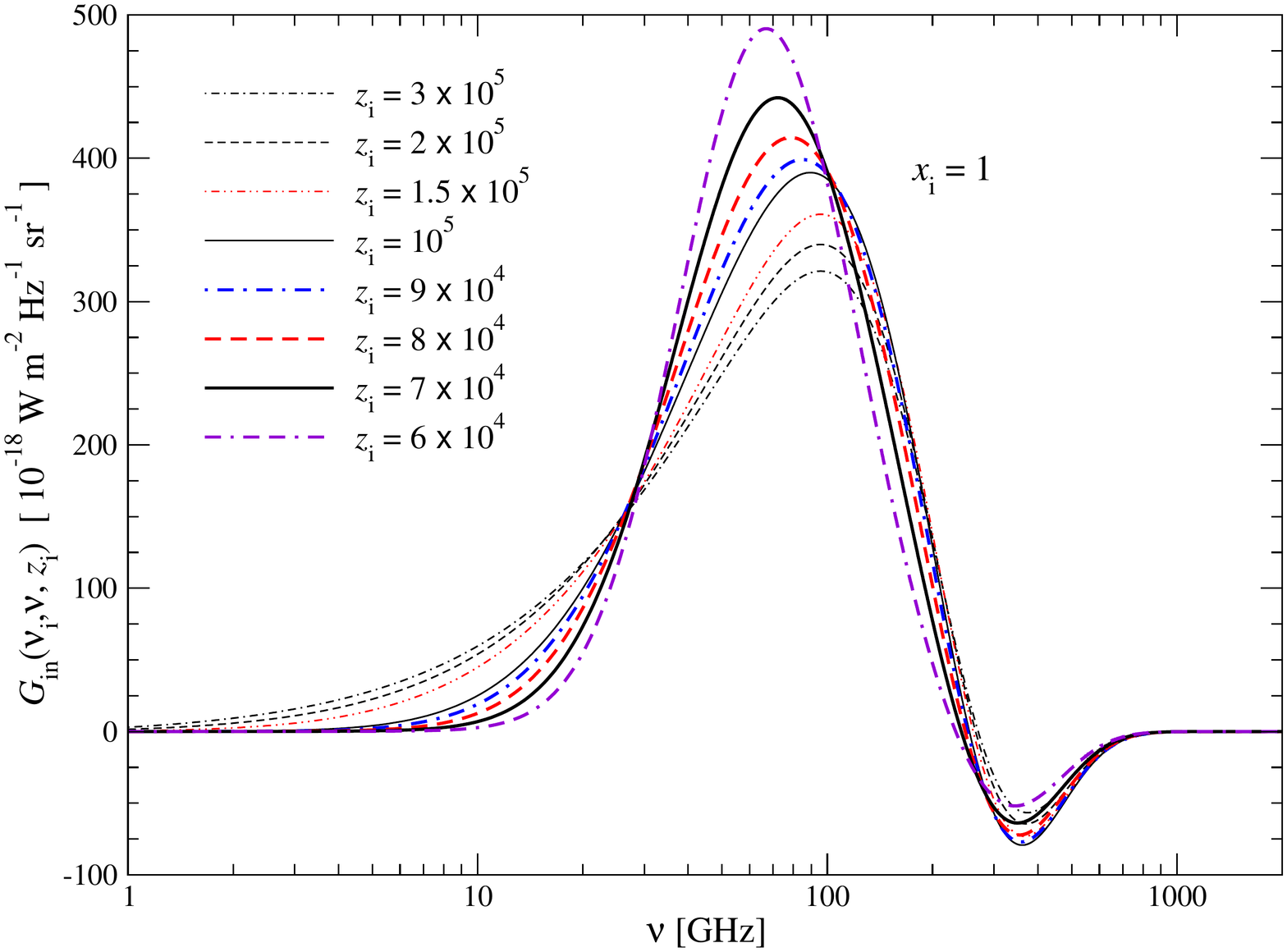}
\\[2mm]
\includegraphics[width=1.04\columnwidth]{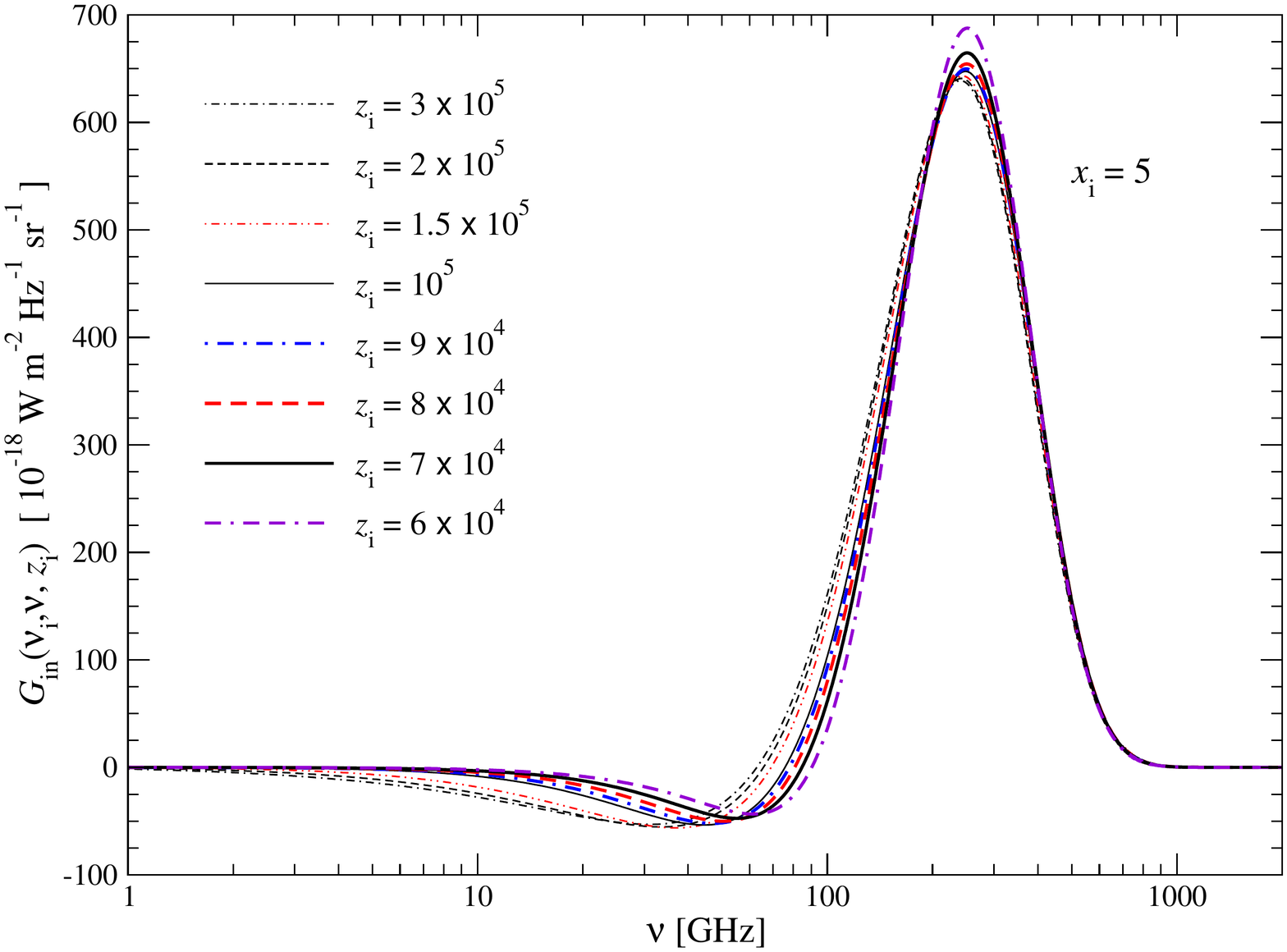}
\includegraphics[width=1.04\columnwidth]{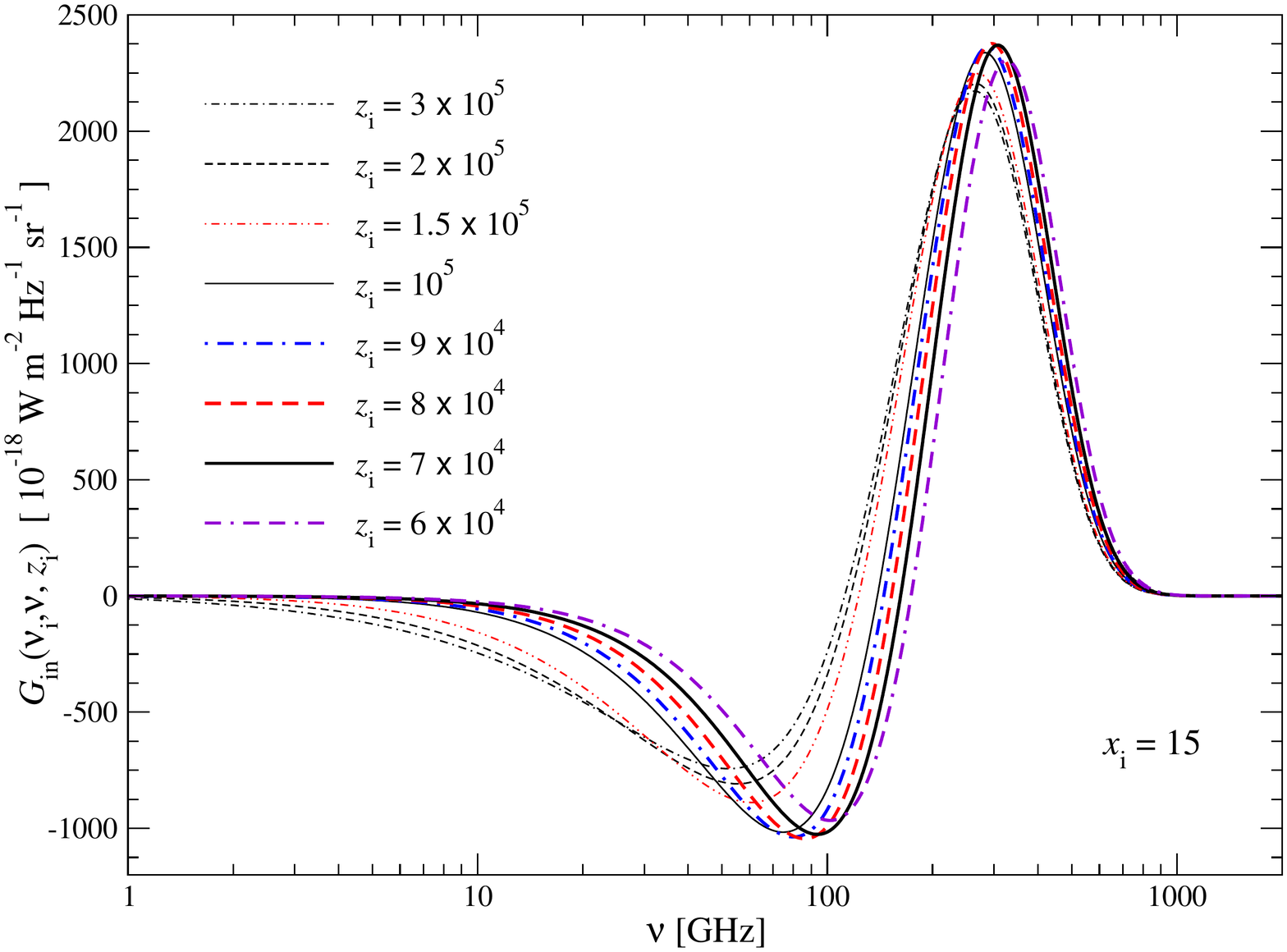}
\caption{Photon injection Green's function for injection at intermediate redshifts, $\pot{5}{4}\lesssim \zin\lesssim \pot{3}{5}$. The photon injection Green's function shows a rich phenomenology. To convert between $x$ and observing frequency today we may use $x\simeq 0.017 \,(\nu/\GHz)$.}
\label{fig:Sols_high}
\end{figure*}

We compared the numerical solution from simple diffusion calculations with this approximation and found that for larger values of $y$ and $\xin$, the position of the line was too low and the width a bit too large. Replacing the dispersion of the Gaussian by $y\rightarrow y/(1+\xin y)$ reproduced the width extremely well, even for larger values of $y$ and $\xin$. 
The match in the position of the line was further improved by replacing $-3y\rightarrow-3y/\sqrt{1+\xin y}$. To improve the match for $\xin\simeq 1$, we need to transition from $-3y\rightarrow-y$ around $\xin\simeq 1$. After several attempts, we found $-3y\rightarrow -y[3-2f(\xin)]$ with $f(\xin)=\expf{-\xin}(1+\xin^2/2)$ to work very well. The match for the dispersion of the line was further improved by replacing $y\rightarrow y/[1+\xin y (1-f(\xin))]$. 

These considerations lead to the refined scattering solution
\beal
\label{eq:Dn_sol_ZS_improved_II}
\Delta n^*(x, y)&=
A \,\frac{\exp\left(-\left[\ln(x/\xin)-\alpha \,y+\ln(1+\xin y)\right]^2/4y \,\beta \right)}{\sqrt{4\pi y \,\beta}\,x^3},
\end{align}
with $\alpha=[3-2 f(\xin)]/\sqrt{1+\xin y}$ and $\beta=(1+\xin y[1-f(\xin)])^{-1}$. The average energy density of the injected photons thus scales as $\Delta\rho_\gamma(y_\gamma)/\rho_\gamma=(\Delta\rho_\gamma/\rho_\gamma)\,\expf{y_\gamma(\alpha+\beta)}/(1+\xin y_\gamma)$. 
Following similar arguments as above, for $x\gtrsim 1$ we then find
\beal
\label{eq:G_x_z_1e3_1e4_high}
G_{\rm in}(\nu, \nu', z)
&\approx 
\left[\frac{c\rho_\gamma(T_0)}{4\pi}\,
\frac{\expf{-\tau_{\rm ff}(x', z)}}{\sqrt{4\pi y_\gamma\beta}\,x'}\,
\expf{-\left[\ln(x/x')-\alpha y_\gamma+\ln(1+x' y_\gamma)\right]^2/4y_\gamma\beta}\right.
\nonumber\\[1mm]
&\qquad+\left.\left(1-\frac{\expf{4y_\gamma(\alpha+\beta)}\expf{-\tau_{\rm ff}(x', z)}}{1+x' y_\gamma}\right)\frac{Y(\nu)}{4}\right] x' \alpha_\rho,
\end{align}
where $\alpha$ and $\beta$ are evaluated at $x'$ and $y_\gamma(z)$. A comparison with the numerical results for $\xin=5$ and several injection redshifts is shown in Fig.~\ref{fig:sol_ill}. Clearly, the new approximation represents the full numerical results very well. 

Over a wider range of injection energies, Eq.~\eqref{eq:G_x_z_1e3_1e4_high} works very well until $\zin \simeq \pot{3}{5}$ (see Fig.~\ref{fig:Sols_low}). For $\xin\simeq 1-5$ we found this solution to work even better, reaching up to $\zin\simeq \pot{5}{4}$. 
At high frequencies, photon absorption is already negligible and we can see from Fig.~\ref{fig:Sols_low} that the net heating/cooling, which gives rise to a $y$-type contribution, can usually be neglected unless we inject at $\xin\gtrsim 1/y_\gamma$, for which recoil becomes significant. 

At $\zin \gtrsim \pot{3}{4}$, the evolved line (omitting the $y$-part) no longer is well approximated by a simple Gaussian, with third moments becoming important (see Fig.~\ref{fig:Sols_low}). Improved approximations that include higher order moments and frequency-dependent dispersion terms may be possible, but we leave this question to future work. In addition, closer to $\zin\simeq \pot{5}{4}$, corrections to the $y$-type contribution due to the $r$-type (non-$\mu$/non-$y$) distortion become significant. This could be captured by computing the effective heating rate from the evolution of the line as a function of redshift and then feeding it into the thermalization Green's function of energy release to threat the heating contribution more precisely.

\subsection{The $\mu$-$y$ transition era}
\label{sec:hybrid}
The signatures of photon injection during the $\mu$-$y$ transition era ($10^4\lesssim z\lesssim\pot{3}{5}$) show the richest phenomenology. In this regime, direct information about the initial distribution of photons can in principle be regained, since Comptonization is no longer able to smear photons out over the whole CMB energy spectrum like during the $\mu$-era. This is also the regime where heating of the matter by the injected photons becomes incomplete, so that the distortion starts to be dominated by the evolution of the injected photons when approaching $\zin\rightarrow 10^4$ and later (Sect.~\ref{sec:y_estimate}).

\begin{figure}
\centering
\includegraphics[width=\columnwidth]{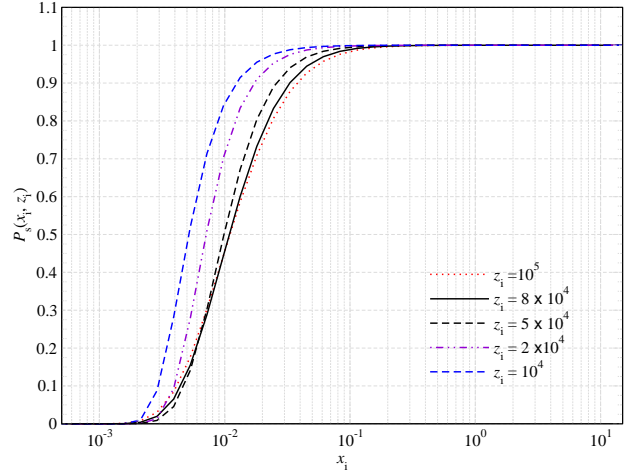}
\caption{Survival probability for different injection frequencies and redshifts after the $\mu$-era. The curves were computed using {\tt CosmoTherm}. At low redshifts ($\zin\lesssim \pot{5}{4}$), Compton scattering becomes inefficient, so that the survival probability steepens from $\Jsurv\approx \expf{-\xc/x}$ to $\approx \expf{-(\xc^*/x)^2}$, where $\xc^*$ can be deduced from Eq.~\eqref{eq:P_BR}.}
\label{fig:P_surv_low}
\end{figure}

In Fig.~\ref{fig:Sols_low} and \ref{fig:Sols_high}, we illustrate the numerical results for several cases. In particular, for injection at high frequencies ($\xin\gtrsim 1$) and $\zin\gtrsim \pot{5}{4}$, the distortion shows large similarities with the distortions from pure energy release. However, due to the addition of photons, the Green's function for photon injection has a significant contribution $\propto G(\nu)$, especially when $\zin\rightarrow \pot{3}{5}$. This is because the injected photons are smeared out over the whole CMB frequency range via Compton scattering without being strongly attenuated by photon absorption and a photon survival probability close to unity (see Fig.~\ref{fig:P_surv_low}). We can also see that for $\xin=1$, a negative $y$-type contribution arises because on average the plasma cools while smearing the injected photons out over the CMB spectrum. We find the transition between net heating and net cooling to occur at $\xin\simeq 3.6-3.83$, depending on the injection redshift (see Fig.~\ref{fig:Regimes_II}).

For injection at lower frequencies, in Fig.~\ref{fig:Sols_high} one can still directly identify the broadened and partially up-scattered photon line until the $y$-parameter exceeds unity significantly ($\zin\gtrsim \pot{2}{5}$). This is because low-frequency photons have to Comptonize significantly until reaching the maximum of the CMB spectrum, a process that requires many scatterings. Focusing on the high-frequency distortion, for $\xin=10^{-3}$, one can also see the transition from net heating to net cooling, which occurs around $\zin \simeq \pot{2}{5}$ (see Fig.~\ref{fig:Regimes_II}). 

In summary, the signals created by photon injection show a richer phenomenology than those caused by single energy release, in particular at $\zin\lesssim\pot{3}{5}$, where the final spectrum is found in a partially Comptonized state. If photons are injected at several frequencies, a superposition of different distortion shapes can leave even richer signatures in the CMB spectrum. However, this also makes it harder to interpret the constraints on individual scenarios in a model-independent way, a problem that will be considered more carefully in a subsequent paper.

\vspace{-4mm}
\section{Photon injection at high energies}
\label{sec:gamma_injection}
The discussion of the preceding sections was limited to photon injection at $\xin\lesssim 30$. Here, we consider injection at higher energies and $\zin\gtrsim 10^3$. For energies below the pair creation threshold with a soft background photon, $\xp\approx2\me c^2/k\Tg\simeq \pot{4.3}{9} / (1+z)$, the injected photons mainly transfer their energy to the medium via electron recoil. In the expanding Universe, we have $\xin(y)=\xin/(1+\xin y)$, so that the injected photon energy density is roughly given by 
\beal
\label{eq:high_gamma_evol}
\frac{\Delta \rho_\gamma(y)}{\rho_\gamma}\approx \frac{\alpha_\rho \xin}{1+\xin y}\frac{\Delta N_\gamma}{N_\gamma}.
\end{align}
This approximation neglects any line broadening though electron recoil and Doppler terms, which are discussed in \cite{Sazonov2000}, but this should only lead to a correction. Equation~\eqref{eq:high_gamma_evol} implies an energy release history
\beal
\label{eq:high_gamma_Q}
\frac{\id(Q/\rho_\gamma)}{\id z}\approx \frac{\alpha_\rho \xin^2 }{(1+\xin y)^2}\frac{\Delta N_\gamma}{N_\gamma}
\,\frac{k\Tg}{\me c^2} \frac{\sigT \Ne c}{H (1+z)},
\end{align}
which can be directly used with the energy release Green's function to compute the distortion signal. Depending on the injection epoch, eventually most of this injected photon energy heats the plasma, causing a $\mu$-, $y$- and $r$-type distortion, with total energy release $\Delta \rho_\gamma/\rho_\gamma\simeq \alpha_\rho \xin \,\Delta N_\gamma/N_\gamma$.
At early times ($\zin\gtrsim \pot{\rm few}{5}$, where Compton scattering is efficient) or for injection at $1/y_\gamma \ll \xin \lesssim \xp$ (so that electron recoil is strong), the energy release for a single photon injection occurs quasi-instantaneously and can be estimated by adding $\Delta \rho_\gamma/\rho_\gamma\simeq \alpha_\rho \xin \,\Delta N_\gamma/N_\gamma$ at $\zin$ to the plasma. However, at lower redshifts, the shape of the distortion depends more directly on the energy loss rate. In this case, differences between the `on-the-spot' approximation\footnote{This means that the injected photon energy is converted into heat quasi-instantaneously \citep{Padmanabhan2005}.} and the detailed energy release history may be significant, especially when considering electromagnetic cascades above the pair-production threshold.

\subsection{Contributions from primary and secondary photons}
The picture given above is rough. In particular, it assumes that the contributions from secondary photons and non-thermal electrons, produced while the hard photons lose their energy, are negligible. Similarly, the injected primary photon spectrum is neglected. 
This last statement is valid simply because for $\xin\gg 1$ and given $\Delta \rho_\gamma/\rho_\gamma$, below the COBE/FIRAS limit, $\Delta \rho_\gamma/\rho_\gamma\lesssim \pot{6}{-5}$ \citep{Fixsen1996, Fixsen2009}, the added primary photon number is $\Delta N_\gamma/N_\gamma=\alpha_\rho^{-1}\,\xin^{-1}\Delta \rho_\gamma/\rho_\gamma\ll \Delta \rho_\gamma/\rho_\gamma$. Thus, only when $N\simeq \xin$ secondary photons are produced, does one have to worry about more than just the heating caused by the injected photon.

One way for the primary photon to produce secondary photons is through DC emission with thermal electrons. Some of these DC photons are directly injected close to the CMB bands, $x\lesssim 30$. To estimate by how much this could change the final spectral distortion, we shall use the DC emissivity in the {\it soft photon limit}\footnote{In this limit, the emitted DC photon has a much lower energy than the scattering high-energy photon.} \citep{Lightman1981, Thorne1981}; however, we add relativistic corrections due to the fact that the injected photon can be hard, $h\nuin\approx \me c^2$. This gives \citep{Chluba2007a}
\bsub
\beal
\label{eq:DC_hard}
\frac{\partial n}{\partial y} 
&\approx  \frac{\Lambda_{\rm h}(x, y)}{x^3\,\Thg}\left[1-n \,(\expf{x \, \phi_{\rm c}(y)}-1)\right],
\\
\Lambda_{\rm h}&\approx \frac{4\alpha}{3\pi}\,\Thg^2 \int \frac{x^4 n_{\rm h}(x)\id x}{1+\frac{21}{5}x\Thg+\frac{84}{25}x^2\Thg^2-\frac{2041}{875}x^3\Thg^3+\frac{9663}{4375}x^4\Thg^4}.
\end{align}
\esub
We neglected stimulated DC emission due to the scattered photon $\nu'\simeq \nuin$, which should cause a tiny correction. We furthermore defined the temperature ratio, $\phi_{\rm c}(y)=\Tg/T_{\rm c}(y)$, where $T_{\rm c}(y)$ is the color temperature of the high-frequency photon, which initially exceeds the CMB photon temperature by a large amount. 

Inserting $n_{\rm h}(x)=G_2^{\rm pl} (\Delta N_\gamma/N_\gamma) x^{-2} \delta[x-\xin(y)]$ for the distribution of the hard photon, with $x_{\rm h}=\xin/(1+\xin y)$, we have the emission coefficient
\beal
\label{eq:DC_Lambda_hard}
\Lambda_{\rm h}&\approx \frac{4\alpha}{3\pi}\,\Thg^2 \frac{G_2^{\rm pl}\,x_{\rm h}^2\,(\Delta N_\gamma/N_\gamma)}{1+\frac{21}{5}x_{\rm h}\Thg+\frac{84}{25}x_{\rm h}^2\Thg^2-\frac{2041}{875}x_{\rm h}^3\Thg^3+\frac{9663}{4375}x_{\rm h}^4\Thg^4}.
\end{align}
For the CMB blackbody photons, most of the DC emission is produced at $x=(\mathcal{I}_{\rm dc}/G_2^{\rm pl})^{1/2}\simeq 3.3$, with the effective DC integral 
$\mathcal{I}_{\rm dc}=\int x^4 n_{\rm b}(1+n_{\rm bb})\approx 25.976$. Thus, the relative efficiency of DC emission caused by the hard photons is 
\beal
\label{eq:DC_Lambda_hard}
\frac{\Lambda_{\rm h}}{\Lambda_{\rm DC}}&\approx \frac{(x_{\rm h}/3.3)^2 (\Delta N_\gamma/N_\gamma)}{1+\frac{21}{5}x_{\rm h}\Thg+\frac{84}{25}x_{\rm h}^2\Thg^2-\frac{2041}{875}x_{\rm h}^3\Thg^3+\frac{9663}{4375}x_{\rm h}^4\Thg^4}.
\end{align}
We find that even if we saturate the COBE/FIRAS bound, $\Delta N_\gamma/N_\gamma\lesssim \pot{6}{-5}/(\alpha_\rho\,\xin)\approx \pot{1.6}{-4}\xin^{-1}$, for $\xin\lesssim \xp$, this extra emission relative to the CMB blackbody DC emission can only be important at\footnote{Neglecting the suppression of DC emission due to relativistic corrections, one obtains $\zin\lesssim \pot{7}{4}$.} $\zin\lesssim 4000$. Since at these low redshifts, DC is already much slower than BR, this only adds a small correction. This statement also holds when going beyond the soft photon limit \citep{Gould1984, Chluba2005}, which allows including the DC emission at higher frequencies.

How many secondary photons could maximally be produced? For $\xin$ below the pair-production threshold, the hard photons will indeed lose their energy mainly through recoil \citep{Zdziarski1989}. At high redshifts ($z\gtrsim 3300$), the Thomson optical depth, $\tau\approx 0.21 (1+z)$, strongly exceeds unity, showing that photons undergo many scatterings. The loss of energy by recoil in a single scattering event ($\tau\simeq 1$) is $\Delta \nu/\nu\simeq - h\nu/\me c^2$. Thus, as long as $h\nuin\ll \me c^2$, the scattered thermal electron only receives a small kick, subsequently sharing its energy with the thermal plasma on a short time-scale. In this case, Eq.~\eqref{eq:high_gamma_Q} describes the situation very well, since hardly any energy is converted into energetic secondary particles that could also enhance the BR and DC emissivities.

For $h\nuin\gtrsim \me c^2$, a significant population of secondary, non-thermal electrons can be built up. These cool down by inverse Compton scattering with CMB photons, which in addition removes photons from the CMB bands, creating additional hard photons. This is equivalent to a photon destruction event (negative photon injection), which could leave a signature in the final distortion. However, the total number of all energetic secondary particles (including up-scattered CMB photons) can never become dramatic due to energetic constraints. The only way to significantly enhance the number of photons may be by very soft photon production ($x\ll 1$), for example, through non-thermal BR. This does not require a lot of energy but could directly help in the thermalization process.

To include these aspects consistently, more detailed cascade calculations in the pre-recombination era are required. These should not only address how the medium is heated as a function of time, an aspect that determines the exact distortion shape ($\mu$-, $y$- and $r$-distortion) caused by heating, but also how many secondary photons are eventually added to or removed from the CMB bands and when. At $\zin\lesssim \pot{\rm few}{5}$, the final secondary photon population reaching the CMB bands from the high energies may also leave its features. Combining these aspects could allow discerning between different scenarios related to decaying or annihilating particles.

\section{Current constraints on photon injection}
\label{sec:FIRAS_const}
\begin{figure}
\centering 
\includegraphics[width=1.03\columnwidth]{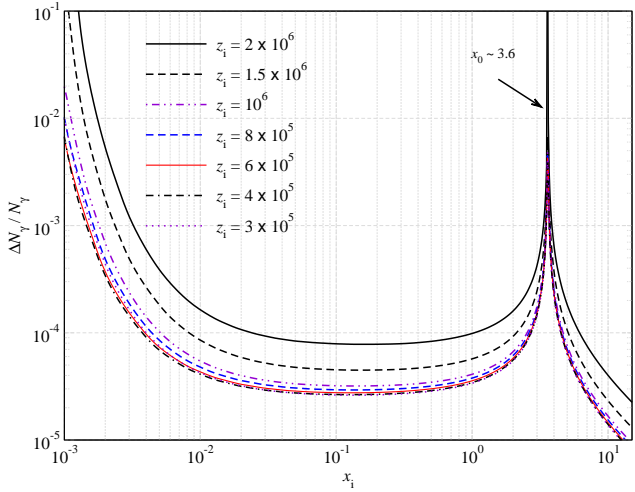}
\caption{COBE/FIRAS limits (68\% c.l.) on photon production for redshifts during the $\mu$-distortion era. Around $\xin\simeq 3.6$, the net chemical potential created by photon production vanishes so that the tightest limit is $\left|\Delta N_\gamma/N_\gamma\right|\lesssim 0.07 \,(68\%\,\text{c.l.})$ from measurements of the light element abundances \citep{Simha2008}.}
\label{fig:mu_limits}
\end{figure}

In this section, we give a brief discussion of current constraints on photon injection from COBE/FIRAS \citep{Fixsen1996, Fixsen2009} and BBN \citep{Simha2008, Jeong2014}. From \cite{Simha2008}, we have 
\beal
\frac{\Delta N_\gamma}{N_\gamma}\simeq -0.08 \pm 0.07\;(68\%\,\text{c.l.})
\end{align}
a number that is derived by comparing the CMB temperature at recombination and BBN, where the latter is derived using measured light element abundances, implying $\left|\Delta N_\gamma/N_\gamma\right|\lesssim 0.07 \,(68\%\,\text{c.l.})$. While this limit is not very tight, it supersedes the constraint from $\mu\lesssim \pot{9}{-5}\,(95\%\,\text{c.l.})$ \citep{Fixsen1996} for photon injection at very low frequencies and around $\xin\simeq 3.6$ (see Fig.~\ref{fig:mu_limits}). At high frequencies, one also has 
\beal
\left|\frac{\Delta N_\gamma}{N_\gamma}\right|\lesssim \frac{\pot{8.9}{-5}}{(\xin-3.6)\,\mathcal{J}^\ast_{\rm bb}} \;(68\%\,\text{c.l.}),
\end{align}
which imposes a very tight limit on energetic photon production during the $\mu$-era.

To derive constraints on photon injection at different frequencies and for $\zin\lesssim \pot{3}{5}$, one would need to consider the precise shape of the final distortion in light of COBE/FIRAS data, since the distortion can no longer be parametrized as a simple $\mu$ or $y$-distortion (see Fig.~\ref{fig:Sols_low} and \ref{fig:Sols_high}). In particular, for $1\lesssim \xin \lesssim 10$, corresponding roughly to the COBE/FIRAS channels $68\,\GHz\lesssim \nu\lesssim 640\,\GHz$, one could expect residual narrow features from the direct photon injection event at these redshifts. While we leave a detailed discussion for the future, for injection at very high and very low frequencies (outside the regime directly probed with COBE/FIRAS) one can still obtain some constraints. For injection at low frequencies ($\xin\ll 1$), the overall heating of the plasma (which would leave a high-frequency $\mu$/$y$-distortion) remains small unless we allow for a very large photon production, $\Delta N_\gamma/N_\gamma\simeq 1$. In this case, more detailed computations of the distortion may be required, since non-linear effects could become important. This statement may also apply to soft photon injection at high redshift.

For injection of energetic photons ($\xin\gtrsim 10$) at $\zin\lesssim \pot{3}{5}$, from Eq.~\eqref{eq:high_gamma_evol} we have the total energy release
\beal
\label{eq:high_gamma_evol_tot}
\frac{\Delta \rho_\gamma}{\rho_\gamma}\approx \frac{\alpha_\rho \xin^2 y_\gamma}{1+\xin y_\gamma}\frac{\Delta N_\gamma}{N_\gamma},
\end{align}
between $\zin$ and today. For photon injection at late times ($10^3\lesssim \zin\lesssim\pot{5}{4}$ and $y_\gamma\ll 1$), this gives 
\beal
\left|\frac{\Delta N_\gamma}{N_\gamma}\right|\lesssim \frac{\pot{8.1}{-5}}{\xin^2 y_\gamma} \,(68\%\,\text{c.l.}).
\end{align}
This limit weakens strongly towards lower redshifts, since only a small fraction $\propto \xin y_\gamma$ of the total injected photon energy is converted into heating of the medium, simply because electron recoil becomes inefficient.

\begin{figure}
\centering 
\includegraphics[width=\columnwidth]{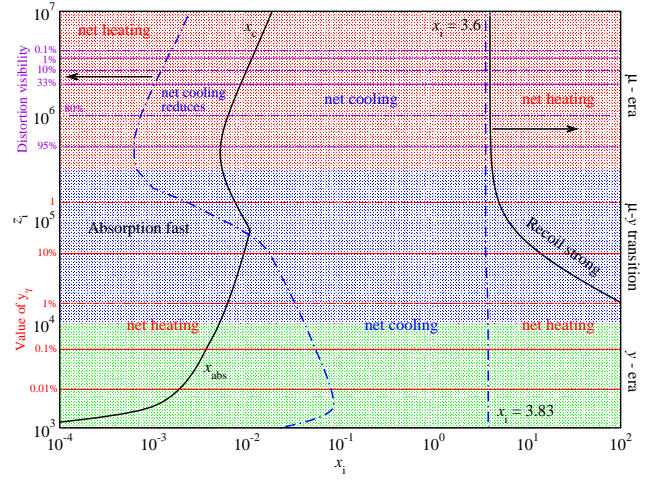}
\caption{Regions for net heating and cooling by photon injection. At high frequencies, depending on the injection redshift, the transition between net cooling and heating occurs around $\xin \simeq 3.6-3.83$, while at low frequencies it occurs over a wide range of frequencies at different redshifts (blue dash-dotted lines). In black, we also show the injection frequency for which the total recoil exceeds unity and the critical frequency at which $\Jsurv\approx 1/{\rm e}$ (optical depth $\tau\approx 1$). The horizontal lines also indicate redshifts of constant $y_\gamma$ (dashed red) and constant distortion visibility (dashed purple).}
\label{fig:Regimes_II}
\end{figure}

\vspace{-4mm}
\section{Conclusions}
\label{sec:conclusions}
We studied the spectral distortions signatures caused by photon injection over a wide range of energies and redshifts. Our calculations illustrate the rich phenomenology of the final distortion shapes, with several examples shown in Fig.~\ref{fig:Sols_low} and \ref{fig:Sols_high}. They also constitute the starting point for future investigations of distortions created by specific photon injection scenarios.

In Sect.~\ref{sec:Greens_nu}, we discussed a detailed picture for the shape of the photon injection Green's function in different regimes. Figure~\ref{fig:Regimes_II} summarizes the different domains. Generally, we find that for photon injection at high ($\xin > 3.6-3.83$) and very low frequencies ($\xin\lesssim 10^{-4}-10^{-3}$) a net heating of the plasma occurs. At low frequencies this is because photons are efficiently absorbed and thus directly converted into heat, while at high frequencies it is because on average photons have to down-scatter to thermalize. At intermediate regimes, a net cooling of the plasma occurs.

For photon injection during the $\mu$-era ($z\gtrsim \pot{3}{5}$), simple analytic approximations for the amplitude of $\mu$ are given in Sect.~\ref{sec:mu_estimate}. Similarly, for moderate scattering $y$-parameter ($\zin \lesssim \pot{3}{4}$), we found simple approximations for the photon injection Green's function. In particular, we obtained a new Compton scattering solution, Eq.~\eqref{eq:Dn_sol_ZS_improved_II} for single photon injection, which is valid over a wide range of energies. The solution merges the high- and low-frequency domain, simultaneously including Doppler broadening and boosting, electron recoil and stimulated scattering. At intermediate redshifts, detailed numerical calculations are required, since the final distortion can be found in a partially Comptonized state (Fig.~\ref{fig:Sols_low} and \ref{fig:Sols_high}). The introduced Green's function method provides a simple way to accelerate the computations in this regime and will be made available at \url{www.Chluba.de/CosmoTherm}.

Our calculations show that photon injection can offset the distortion signals created by pure energy release. In particular, since photon injection can create negative $\mu$ and $y$ distortion contributions, this may lead to significant limitations for the interpretation of future CMB spectral distortions measurements. However, for photon injection at $\zin\lesssim\pot{3}{5}$, partial direct information about the photon injection process may be recovered. This could help distinguishing pure energy release and photon injection scenarios, a problem that will be considered in a future analysis.

We briefly discussed photon injection at high energies, reaching close to and above the pair-production threshold (Sect.~\ref{sec:gamma_injection}). Our analysis indicates that above the pair-production threshold, more detailed computations of the soft photon production efficiency ($x\lesssim 1$) may be required. These soft photons could help thermalizing the distortion and also may introduce features into the final signal that could further help discerning different processes. Similarly, detailed considerations of the reprocessing of photons and energy during the post-recombination by atomic and molecular species should be considered more carefully.

In Sect.~\ref{sec:FIRAS_const}, we presented COBE/FIRAS and BBN constraints on photon injection. These could be improved significantly with a PIXIE-like experiment. The constraints in the $\mu$-era are summarized in Fig.~\ref{fig:mu_limits}, showing that scenarios with significant photon production around $\xin \simeq 0.01-10$ are already excluded. However, at very low frequencies ($\xin \lesssim 0.01$) and around $\xin\simeq 3.6$, in principle significant photon injection ($\Delta N_\gamma/N_\gamma\simeq 1$) is still allowed and may require the inclusion of non-linear effects. For $\xin\lesssim 0.01$ this is simply because the added energy $\Delta \rho_\gamma/\rho_\gamma\simeq\alpha_\rho \xin \Delta N_\gamma/N_\gamma$ is very small unless $\Delta N_\gamma/N_\gamma$ becomes large, while for $\xin\simeq 3.6$ a balanced photon injection scenario, with vanishing net distortion (see Sect.~\ref{sec:Teq}), is encountered. For photon injection at lower redshifts, a detailed comparison with COBE/FIRAS or future spectral distortion data has to be carried out, since in principle narrow features can remain in the partially Comptonized regime. We leave this to future work.

We close by noting that further strong motivation for considering scenarios with photon injection can be given due to the observed low-frequency excess at $h\nu\simeq 3.3\,\GHz$ \citep{Fixsen2011excess}, which may have an interpretation as a signature of a decaying or annihilating particle in the pre-recombination era at $\zin \lesssim 10^5$. In this case, soft secondary photons could be produced that now are found in a partially Comptonized state, however, more detailed computations for specific injection scenarios are required.

\small
\vspace{-5mm}
\section*{Acknowledgments}
The author is supported by the Royal Society as a Royal Society University Research Fellow at the University of Cambridge, UK. 

\small 

\vspace{-5mm}
\bibliographystyle{mn2e}
\bibliography{Lit}

\end{document}